\documentclass[manuscript]{aastex}

\slugcomment{To appear in AJ}

\shorttitle{Young Cluster Metallicities}
\shortauthors{Monroe \& Pilachowski}

\begin{document}

\title{Metallicities of Young Open Clusters I:  NGC 7160 and NGC 2232}

\author{TalaWanda R. Monroe\altaffilmark{1} and Catherine A. Pilachowski\altaffilmark{1}}
\affil{Department of Astronomy, Indiana University, Bloomington, IN 47405, USA}
\email{tmonroe@astro.indiana.edu}
\email{catyp@astro.indiana.edu}

\altaffiltext{1}{Department of Astronomy, Indiana University, Swain West 319, 727 East Third Street, Bloomington, IN 47405-7105}

\begin{abstract}

We present a moderate-resolution spectroscopic analysis of the 10--25 Myr clusters NGC 7160 and NGC 2232, using observations obtained with the WIYN 3.5-m telescope.  Both NGC 7160 and NGC 2232 are found to have super-solar metallicities, with a mean [Fe/H] = 0.16 $\pm$ 0.03 (s.e.m.) for NGC 7160, and 0.22 $\pm$ 0.09 (s.e.m.) or 0.32 $\pm$ 0.08 for NGC 2232, depending on the adopted temperature scale.  NGC 7160 exhibits solar distributions of Na, Fe-peak, and $\alpha$-elements.  NGC 2232 is underabundant in light elements Al and Si, by $\sim$0.25 and $\sim$ 0.15 dex, respectively; [Ni/Fe] is roughly solar.  The abundance of lithium in NGC 2232 stars is in agreement with undepleted values reported for other cluster main sequence stars.  Our abundances are similar to other metal-rich open clusters and Galactic thin and thick disk stars.

\end{abstract}

\keywords{Galaxy: open clusters and associations: general --- Galaxy: open clusters and associations: individual(NGC 7160, NGC 2232) --- stars: circumstellar matter  --- stars: abundances}

\section{Introduction}

Circumstellar dust has been observed around hundreds of main sequence stars in nearby regions of the Galaxy (e.g., Backman \& Paresce 1993; Rieke et al. 2005; Su et al. 2006; Bryden et al. 2006).  Dust in the later-stage debris disks is generally thought to have been generated during collisions between planetesimals that have been gravitationally stirred by massive planets (e.g., Wyatt 2008).  Interest in the origin and evolution of young solar systems, in particular, has resulted in a trove of infrared Spitzer Space Telescope (Werner et al. 2004) observations of young star clusters and stellar associations (e.g., Hernandez et al. 2006; Currie et al. 2007, 2008a, 2008b, 2009; Carpenter et al. 2009).

Young star clusters are good laboratories for the investigation the formation and evolution of planetary systems, including debris disks, and much effort has been exerted to characterize the frequency of debris disks with stellar age (e.g., Rieke et al. 2005; Su et al. 2006; Hernandez et al. 2006; Currie et al. 2008a, 2008b; Carpenter et al. 2009).  Young clusters have well-determined ages and stars within the clusters share the same chemical composition.  Unfortunately, detailed elemental abundances are not available for many of the clusters observed with the Multiband Imaging Photometer for Spitzer (MIPS; Rieke et al. 2004), and for other clusters, metallicity data may be affected by systematic differences among studies.  Investigations of the radial velocity distributions of candidate members and spectral coverage of the youth-indicator lithium would be of value to ongoing efforts to characterize the frequencies of debris disks in young clusters, especially where memberships of solar-type stars are unknown.

As part of a larger study of the metallicities of star clusters surveyed with Spitzer, we present in this paper the first chemical abundance analyses of two young open clusters, NGC 7160 and NGC 2232, in the $\sim$10--30 Myr age range.  Our analysis utilizes cluster F- and G-type dwarfs for which abundances can be determined systematically on a uniform abundance scale, using a variety of metal lines in the spectra of slowly rotating stars.  Future papers will discuss the relationship between age, metallicity, and the frequency of young debris disks.

\section{Cluster and Target Star Selection}

The clusters included in this study are selected from those observed by Spitzer-MIPS at 24 $\micron$.  Open clusters were chosen from the Spitzer Reserved Observations Catalog, satisfying the criteria of youth (each cluster was $<$200 Myr in age) and proximity ((m-M)$_{V}$$<$11.5).  The latter condition was imposed for two reasons:  the sensitivity of MIPS at 24 microns offers assurance that debris disk candidates can be detected around early-type stars in these clusters, and also that solar-type members are typically brighter than V$\sim$16, so that spectroscopic observations can be obtained with 4-m class telescopes.   A brief discussion of the literature and ages of the clusters presented in this paper is given here, along with a detailed description of the sample selections.  The remaining clusters will be discussed in future papers.

\subsection{NGC 7160}

NGC 7160 is the older of two young clusters in the Cepheus OB2 region (l=104.0$\degr$, b=+6.5$\degr$), a bubble of atomic and molecular gas (Simonson \& van Someren Greve 1976; Patel et al. 1994, 1998) located 900 pc away (Sicilia-Aguilar et al. 2005).  NGC 7160 is located in the central region of the bubble, and the younger cluster, Tr 37, is located near the edge of the region.  The age of NGC 7160 was found to be $\sim$10--12 Myr from isochrone fits to the lower main sequence (Sicilia-Aguilar et al. 2004), which is consistent with estimates from dynamics of the gas that suggest the Cep OB2 bubble has an expansion age of 7-10 Myr (Patel et al. 1998).

\begin{figure}
\epsscale{1.00}
\plotone{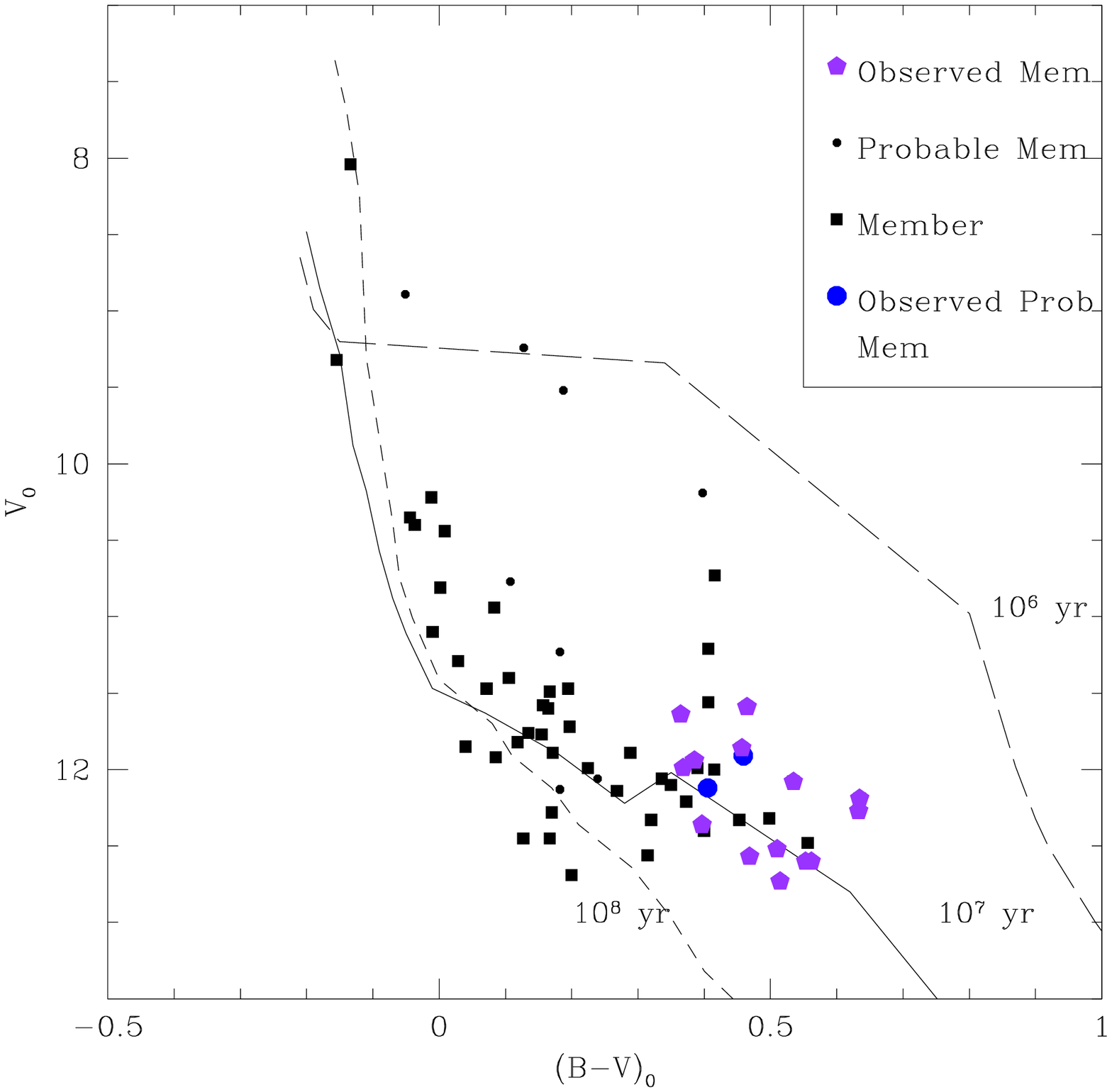}
\caption{A dereddened color-magnitude diagram for stars in NGC 7160, showing the stars observed in this study.  Observed photometry is from Sicilia-Aguilar et al. (2005), which was originally taken from De Graeve (1983).  We dereddened each star using individual A$_{V}$ values from Sicilia-Aguilar et al.  and the relation A$_{V}$=3.17E(B-V), and overplotted 1, 10, 100 Myr isochrones from Siess et al. (2000) (see also Figure 6b in  Sicilia-Aguilar et al.).  Filled squares:  Members according to A$_{V}$ (Sicilia-Aguilar et al.).  Colored pentagons:  Members observed spectroscopically for this study.  Small dots:  Probable members according to A$_{V}$ (Sicilia-Aguilar et al.).  Large colored circles: Probable members observed spectroscopically.}
\label{f1}
\end{figure}

Optical photometric studies include UBV photometry presented by De Graeve (1983) for stars brighter than V$\sim$14, and VRI photometry from Sicilia-Aguilar et al. (2005) for stars fainter than V$\sim$13.5.  Additionally, Sicilia-Aguilar et al. provided spectral types for candidate cluster members in NGC 7160 from low-resolution optical spectra, and also calculated individual extinctions for each star to use as a cluster membership criterion.  They found the average cluster extinction to be A$_{V}$=1.17 $\pm$ 0.45 (s.d.\footnote{Standard deviation.}) mag from 39 candidate members.  The authors attribute the large standard deviation in extinction to uncertainties arising from spectral classification, colors, and photometry, as they found no apparent spatial variations in A$_{V}$ across the cluster.  Stars within 1$\sigma$ of the cluster mean A$_{V}$ were classified as members, and stars having extinctions within 1-2$\sigma$ of the mean were listed as probable members.  We selected F- and G-type stars from their sample (which includes stars from the De Graeve study) that had visual extinctions favorable for membership as program stars.  The stars we selected for observation are presented in a color-magnitude diagram (CMD) in Figure \ref{f1}, along with model isochrones from Siess et al. (2000).  Of the stars presented in our analysis, three are categorized as weak-lined T-Tauri stars (WTTS).  These WTTS stars are not included in Figure \ref{f1} due to the lack of available B photometry.

\subsection{NGC 2232}

NGC 2232 is a sparse and relatively poorly-studied young cluster in the Gould Belt (l=214.6$\degr$, b=-7.4$\degr$).  A photometric UBV study of the brightest candidate members of the cluster was carried out by Claria et al. (1972).  The only optical CCD photometric study presenting UBVRI observations was carried out by Lyra et al. (2006) for the inner 14' x 14' region of the cluster.  Lyra et al. used several pre-main-sequence isochrones to constrain the nuclear age of the cluster to $\sim$30 Myr, and the contraction age to $\sim$25-30 Myr, with the best fit corresponding to 25 Myr.  Lyra et al. also determined the cluster distance of about 320 pc, and interstellar reddening of E(B-V)=0.07.

\begin{figure}
\epsscale{1.00}
\plotone{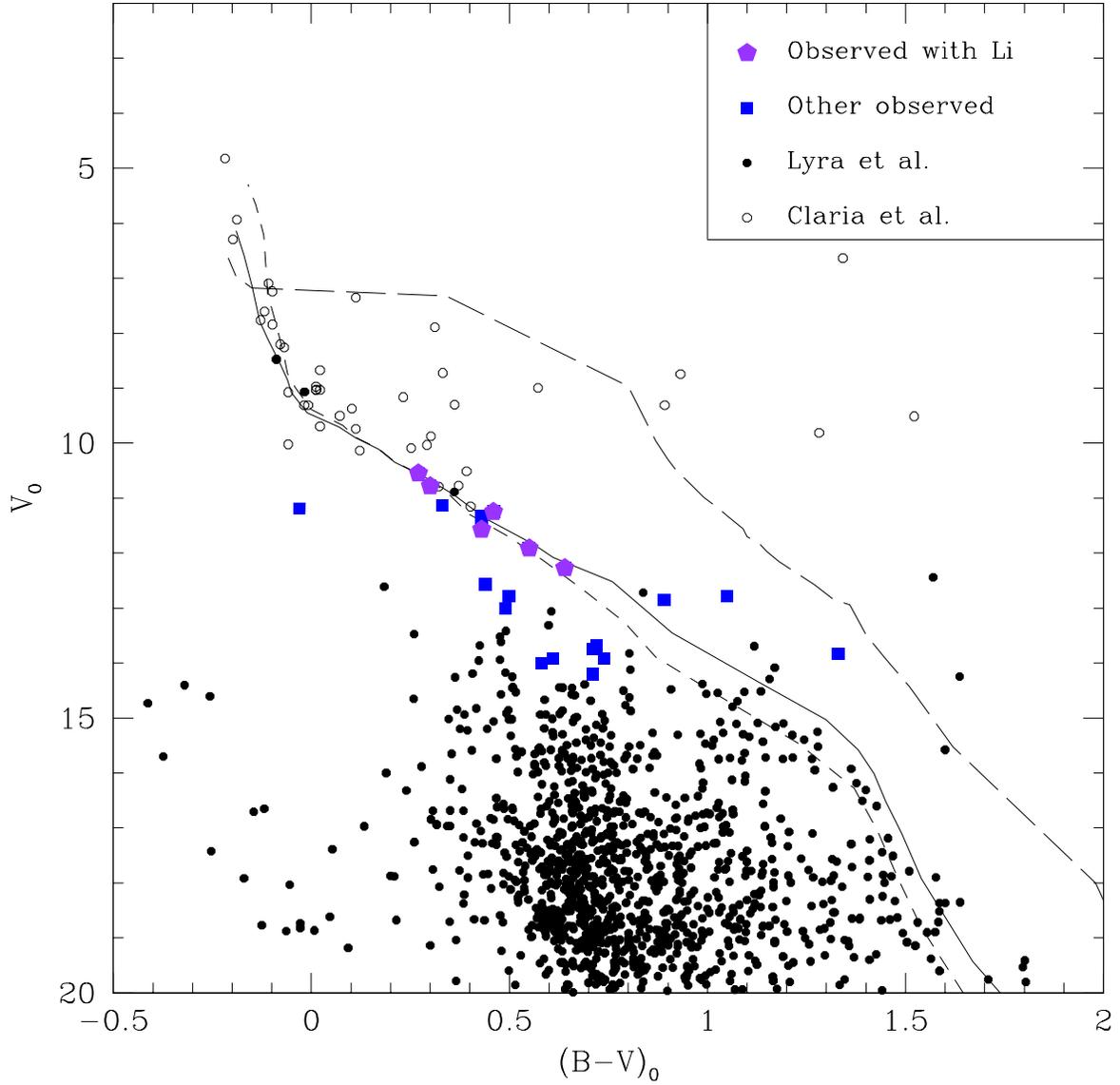}
\caption{A dereddened color-magnitude diagram for observed stars in NGC 2232.  Observed photometry is from Lyra et al. (2006; filled dots) and Claria et al. (1972; open circles).  The Claria et al. photometry has been adjusted to the scale of Lyra et al. by the offsets presented in their Table 4 ($\Delta$V = 0.014 and $\Delta$(B-V) = 0.032).  The data were extinction corrected and dereddened according to the usual relation,  A$_{V}$=3.17E(B-V), and the average cluster reddening of E(B-V)=0.07 (Lyra et al.).  Overplotted are 1, 25, 100 Myr isochrones (long dash, solid line, and short dash, respectively) from Siess et al. (2000).  Filled squares:  Stars observed spectroscopically in the cluster field.  Colored pentagons:  Observed single stars with Li absorption lines.}
\label{f2}
\end{figure}

Currie et al. (2008b) present membership criteria for NGC 2232, including ROSAT X-Ray observations of sources in the cluster field, and spectral types from low resolution spectra for a few dozen candidate members.  Photometric F- and G-type candidate members with (B-V) colors and magnitudes consistent with a 25 Myr isochrone, were chosen for observation based on the CMD from Lyra et al. (2006).  The stars we observed are presented in the CMD in Figure \ref{f2}, with model isochrones from Siess et al. (2000).  We also included additional stars in the field in our observations when possible.  Note that we only observed candidate members from the deeper Lyra et al. sample, and not stars from the Claria et al. study.  The Claria et al. photometry in Figure \ref{f2} was obtained from the WEBDA open cluster database\footnote{See http://www.univie.ac.at/webda/}.

\section{Observations}

Program stars were observed with the multi-object spectrograph Hydra on the WIYN 3.5-m telescope on two different observing runs.  Observations of NGC 7160 were obtained in 2006 December at 6125--6435 \AA, while NGC 2232 was observed in 2007 February and 2007 March at 6380-6800 \AA.  The spectrograph setup for both runs included the 200 $\micron$ red-optimized fibers, the 316 line mm$^{-1}$ echelle grating, the red camera, and the T2KA detector.  Only one fiber configuration was observed per cluster.  Observing conditions were generally clear for the three-night 2006 December run, except for 2006 December 05, when two exposures of NGC 7160 were taken through clouds.  Sky conditions were clear for the 2007 February run, and seeing was generally good.  Sequences of 35 to 60 min exposures were taken for each spectrograph configuration (seven hours total for NGC 7160 and six hours total for NGC 2232), to allow removal of cosmic rays.  A detailed list of dates and exposure times for the observing runs is provided in Table $\ref{tab1}$.

\begin{figure}
\epsscale{1.00}
\plotone{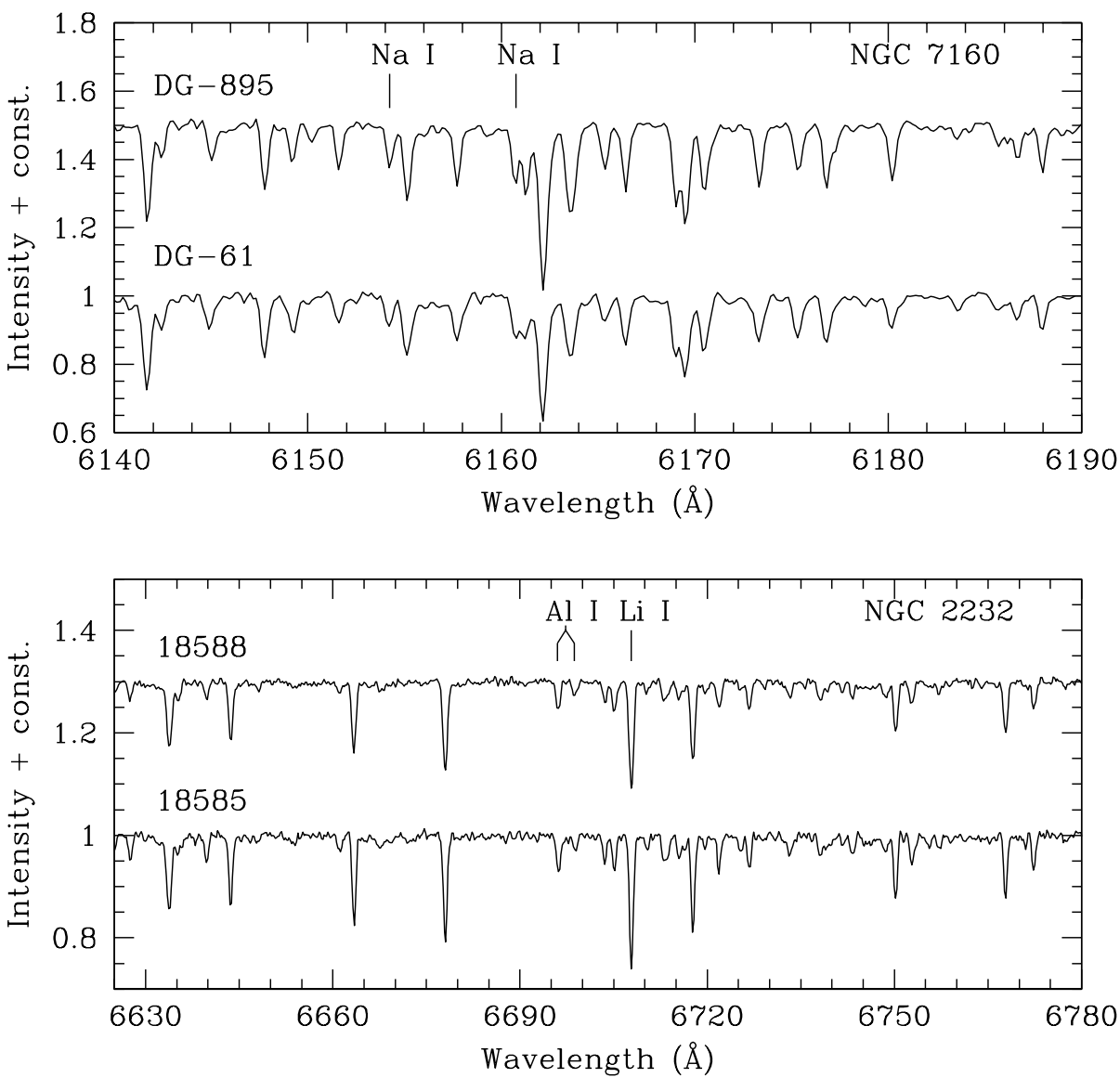}
\caption{Hydra spectra of four representative cluster stars included in our analysis.  Top panel:  Spectra of stars NGC 7160 DG-895 and DG-61 in the region covering the $\lambda$$\lambda$6154,6160 Na doublet.  Bottom panel:  Spectra of NGC 2232 stars 18588 and 18585, which span the region of the $\lambda$$\lambda$6696,6698 Al doublet and $\lambda$6708 Li line.  The continua of stars DG-895 and 18588 have been shifted by 0.5 and 0.3, respectively, for display purposes.  All spectra have been shifted to rest wavelengths.}
\label{f3}
\end{figure}

The data were reduced using standard processing routines in IRAF\footnote{IRAF is distributed by the National Optical Astronomy Observatories, which are operated by the Association of Universities for Research in Astronomy, Inc., under cooperative agreement with the National Science Foundation.}, including second-order bias subtraction, flat field division, extraction of one-dimensional spectra, wavelength dispersion correction, and normalization of the stellar continua.  Prior to extraction of one-dimensional spectra, the images were cleaned of cosmic rays with the IRAF script, ``L. A. Cosmic,"\footnote{See http://www.astro.yale.edu/dokkum/lacosmic/.} (van Dokkum 2001; spectroscopic version).  The fit of the dispersion solution used for the ThAr exposures produced typical residuals of 0.01 \AA.  Individual extracted spectra were then combined to make a single, high signal-to-noise ratio (S/N) spectrum for each star.  Sample spectra of stars in NGC 7160 and NGC 2232 are shown in Figure \ref{f3}. The combined spectra have a resolving power of R$\sim$16,000 and S/N ratios of $\sim$70--300 pixel$^{-1}$.  The regions of continuum that were available to measure the S/N ratios are included in Table \ref{tab2}.

Solar daylight-sky spectra were obtained during each observing run to use as a radial velocity standard and to permit an analysis differentially with respect to the Sun.  Lines contaminated by telluric absorption were omitted from our analysis, since no telluric corrections were made to the spectra.

\section{Radial Velocities \& Membership}
In addition to determining chemical abundances of stars in NGC 7160 and NGC 2232, we present radial velocity studies of the stars observed in both clusters.  Old open clusters typically are characterized by narrow velocity dispersions that make their members kinematically distinct from foreground and background stars in the galactic disk.  As such, radial velocity is often a useful membership indicator.  The use of radial velocity as a membership criterion in young clusters, however, depends on whether the cluster is gravitationally bound.

Radial velocities of stars observed in NGC 7160 and NGC 2232 were determined using the cross-correlation task FXCOR in IRAF.  The object spectrum was cross-correlated with a solar daytime sky spectrum, obtained during the same observing run, as a template.  The resulting cross-correlation peak is fit with a Gaussian and the radial velocity is determined from the location of this peak (Tonry \& Davis 1979).  Cross-correlations were performed on the final combined spectrum for each cluster star, as well as on single, individual exposures to assess the velocity uncertainties.  No fewer than five points of the cross-correlation peak were used for the Gaussian fits and the resulting R values (the peak value of the cross-correlation weighted by the noise) were greater than 11, and typically $\sim$40.  Rotational broadening of the stellar lines in these young stars, particularly early F-stars, resulted in cross-correlations that were not well fitted by Gaussians for a number of candidate members.  For these cases we convolved the solar spectrum with a Gaussian kernel to better match the FWHM of the spectral lines in the object spectra.  The typical RMS scatter between individual exposures for observations of NGC 7160 is $\sim$0.4 km s$^{-1}$.  Inspection of radial velocity measurements of NGC 2232 stars during the two nights of observations revealed an offset of $\sim$1 km s$^{-1}$ between the two nights due to instrumental shifts.  Uncertainties in radial velocities determined for both clusters are less than 2 km s$^{-1}$.

Tables \ref{tab3} and \ref{tab4} contain heliocentric-corrected radial velocities, obtained from the combined spectra, for the NGC 7160 and NGC 2232 fields, respectively.  The errors reported for each star are the standard deviations of radial velocities determined from individual exposures.  Table  \ref{tab3} includes photometry, spectral types, and membership determinations for NGC 7160 from Sicilia-Aguilar et al. (2005).  In addition to the A$_{V}$-selected ``confirmed members" (represented by ``Y") and ``probable members" (represented by ``P") from Sicilia-Aguilar et al., we included additional stars thought to be nonmembers (represented by  ``PN").  Table  \ref{tab4} contains optical photometry for stars in NGC 2232 from Lyra et al. (2006), with star identifications from Currie et al. (2008b).  The observed spectral region for NGC 2232 included the lithium line at $\lambda$6708 \AA, which we use as our main membership criterion.  Stars with Li lines are indicated by ``Y" in Table \ref{tab4}, and stars without Li are indicated by ``N."  Again, extra stars were included in the observations since additional fibers were available in the Hydra configuration.

\begin{figure}
\epsscale{1.00}
\plotone{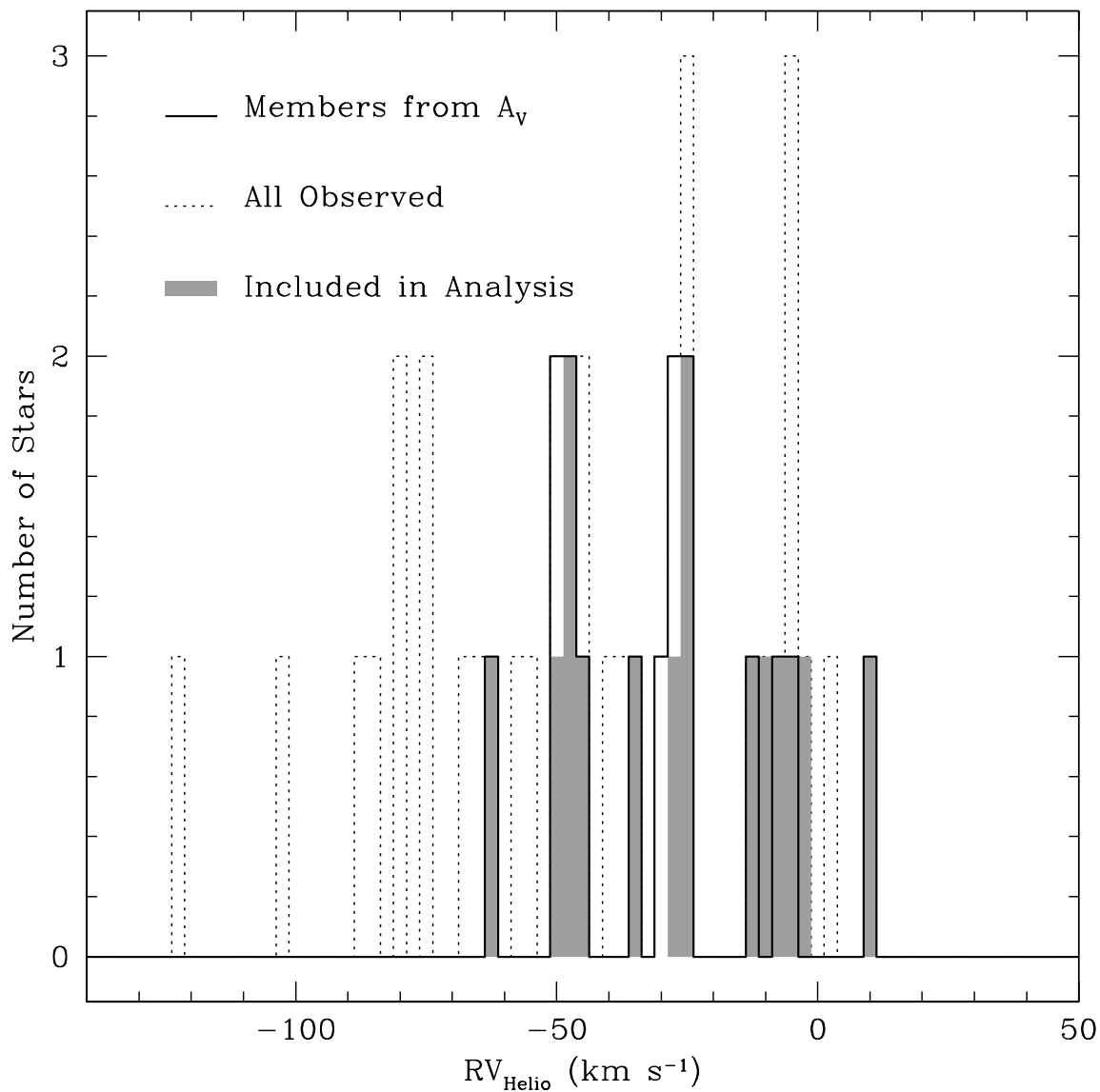}
\caption{A histogram of heliocentric radial velocities for stars in the field of NGC 7160.  All observed stars are indicated by the dotted line.  All observed stars thought to be members based on A$_{V}$ from Sicilia-Aguilar et al. (2005) are indicated by the solid line.  The shaded region of the diagram indicates the stars included in our abundance analysis.  Note that two stars included in the abundance analysis are considered probable members from Sicilia-Aguilar et al.}
\label{f4}
\end{figure}

Figure \ref{f4} shows the velocity distribution of stars in NGC 7160.  Included in the histogram are all stars observed in the field, as well as stars thought to be members by Sicilia-Aguilar et al. (2005), based on visual extinctions.  The stars thought to be members based on A$_{V}$ span $\sim$70 km s$^{-1}$, are centered around -20.4 km s$^{-1}$, and have no modal velocity.  The lack of a peak with a  narrow velocity dispersion in the histogram, which is expected from open clusters in virial equilibrium, suggests that radial velocity may not be a useful membership criterion for NGC 7160.  Since there are no youth-indicator features found in the 6125--6435 \AA\ spectral region, we must rely on the visual extinctions of Sicilia-Aguilar et al. (2005) as our only membership criterion.  The stars selected for our abundance analysis are included in the shaded portion of Figure \ref{f4}.  The large spread in radial velocity suggests that NGC 7160 may be a gravitationally unbound cluster.  The kinematic study by Patel et al. (1998) of the surrounding gas in the adjoining Cepheus OB2 bubble, shows that CO emission extends over a $\sim$35 km s$^{-1}$ range in velocity in the expanding bubble.  

Radial velocity measurements were carried out on early type stars in NGC 7160 by Conti \& van den Heuvel (1970), Liu et al. (1989), and Huang \& Gies (2006).  The stars observed by Conti \& van den Heuvel display a spread in radial velocities, ranging from -36.0 to +2.0 km s$^{-1}$.  Huang \& Gies reported a similar $\sim$50 km s$^{-1}$ spread in radial velocities for B stars in the cluster, which span -50.6 to 2.4 km s$^{-1}$.  Both of these studies estimate their errors to be $\sim$10 km s$^{-1}$.  Liu et al. observed five stars in the cluster, of which two were determined to be spectroscopic binaries based on their large velocity variations.  The remaining three stars (NGC 7160 HG 3, 5, and 6), had velocities of -17.4, -18.8, and -34.6 km s$^{-1}$; from which Liu et al. reported a cluster average of -23.6 $\pm$ 9.6 (s. d.) km s$^{-1}$.  The large spread in radial velocities for the solar type stars in NGC 7160 is consistent with velocity observations of early type stars.

\begin{figure}
\epsscale{1.00}
\plotone{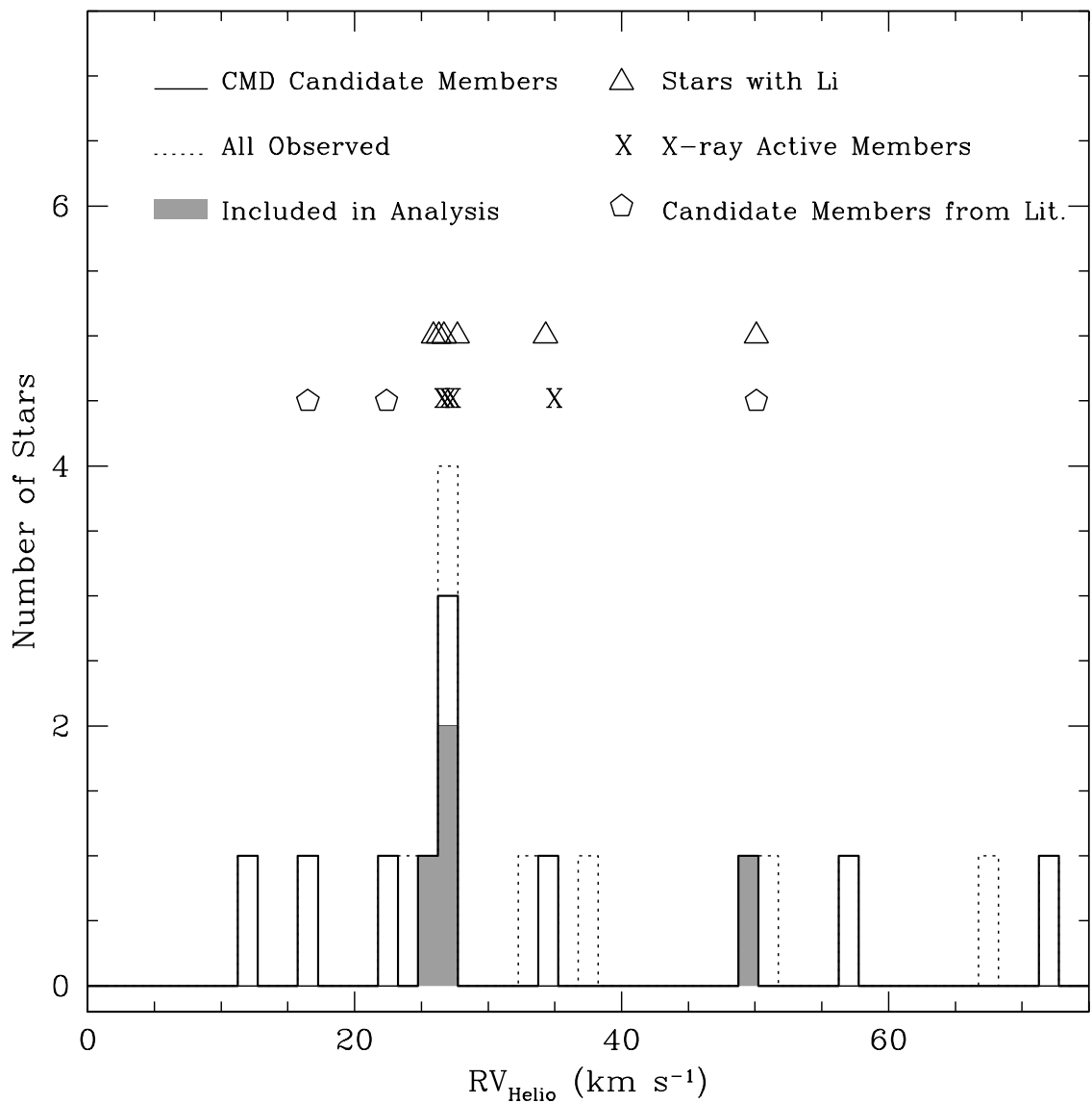}
\caption{A histogram of heliocentric radial velocities for single stars in NGC 2232. All observed stars in this velocity range are indicated by the dotted line.  All observed stars thought to be candidate members from the CMD of Lyra et al. (2006) are indicated by the solid line.  The shaded region of the diagram indicates the stars included in our abundance analysis, including star 9242, whose membership is questionable.  Triangles:  Single stars with Li absorption lines.  Pentagons:  Candidate photometric members from NIR and optical CMDs of Currie et al. (2008b).  Stars marked with ``X" are X-ray active members from Currie et al.}
\label{f5}
\end{figure}

Figure \ref{f5} contains a velocity histogram similar to Figure \ref{f4} for single stars in the NGC 2232 field.  The entire set of stars observed spans a velocity range of -33 to 88 km s$^{-1}$, including the extra stars that were included in the Hydra configuration.  In Figure \ref{f5} we limit the velocity range to include only the stars that were presumed to be photometric members  from Lyra et al. (2006).  Here we also add other membership indicators, including the presence of lithium at $\lambda$6708 \AA,  X-ray activity, and photometric candidacy from the CMD (Currie et al. 2008b).  The six single stars with Li absorption lines span 25.9--50.1 km s$^{-1}$, with four stars near an average velocity of 26.6 $\pm$ 0.77 km s$^{-1}$.  Three of the X-ray active members are at $\sim$26 km s$^{-1}$, while one is at 34.3 km s$^{-1}$.  Based on these membership indicators, we have selected three stars as likely members of NGC 2232 for our abundance analysis, and included star 9242 for analysis as well, until its membership can be further assessed.  These four stars are included in the shaded region of Figure \ref{f5}.

Kharchenko et al. (2007) reported a velocity of 19.85 $\pm$ 4.5 (s.e.m.) from velocity determinations of four early-type stars in NGC 2232 from the literature.  The values reported are 25.3, 10.0, 14.6, and 29.5 km s$^{-1}$ for NGC 2232 1, 2, 3, and 9 (Kharchenko et al. 2007; Gontcharov 2006, and references therein).  Radial velocity data compiled at the WEBDA open cluster database from several sources indicate that stars NGC 2232 1 and 2 are radial velocity variable stars.  

Cross-correlations performed using FXCOR revealed two double lined spectroscopic binary stars in NGC 2232.  The characteristic double cross-correlation peaks were deblended and radial velocities were determined for each star.  Velocity measurements for the seven exposures obtained on the two nights of cluster observations are presented in Table \ref{tab5}.  The heliocentric Julian Date (HJD) is included for the midpoint of each observation.  The stars with the higher cross-correlation peak and smaller cross-correlation peak in each pair were were assigned labels ``A" and ``B," respectively.  The average velocities of the two binary systems are close to the cluster mean at 25.4 and 34.9 km s$^{-1}$ for stars 18650 and 11242, respectively; however, neither binary has a position in the CMD of Figure \ref{f2} that suggests membership in NGC 2232.  The spectra of both binaries contain two sets of absorption lines, and the spectrum of star 18650 has two Li absorption lines present at wavelengths expected for the two stars.  Radial velocities for both components of each binary are included in Table  \ref{tab5}.

\section{Abundance Analysis}

We have determined elemental abundances for the cluster stars using a local thermodynamic equilibrium (LTE), detailed, model atmosphere analysis.  Abundances were derived differentially with respect to the Sun using the LTE line analysis code MOOG (Sneden 1973), and Kurucz (ATLAS9) model atmospheres\footnote{See http://kurucz.harvard.edu/grids.html.}, with no convective overshooting.  

\subsection{Atomic Data and Equivalent Widths}

Our line lists have been compiled from several sources.  Suitable lines were first identified in the solar atlas (Wallace et al. 1998) and Moore's revision of Rowland's solar table of atomic lines (Moore et al. 1966).  Unblended lines for which reliable atomic information could be found were selected for analysis.  The analyses for NGC 7160 and NGC 2232 are discussed below.

NGC 7160:  The 6125--6435 \AA\ spectral region contains $\sim$20 Fe I lines, 4 Fe II lines, and a few lines each of Na, Si, Ca, Sc, Ti, V, Cr, and Ni that could be used for analysis.  Some lines were omitted from our analysis due to the presence of overlapping or nearby diffuse interstellar bands (DIBs), characterized by FWHM $\sim$ 1 \AA.  For Fe I, Fe II, and Ni I we adopt the oscillator strengths of Kurucz \& Bell (1995)\footnote{The atomic line database may be queried at http://www.cfa.harvard.edu/amp/ampdata/kurucz23/sekur.html}.  For the remaining species we rely on log gf values determined from the solar spectrum by Thevenin (1990) or Reddy et al. (2003).  Excitation potentials of the lines were obtained from Moore et al. (1966).

NGC 2232:  In the 6380--6800 \AA\ spectral region we selected $\sim$30 Fe I lines and 3 Fe II lines, which could be measured in the solar daylight sky spectrum.  Of those lines about 20 Fe I lines could be measured in spectra of our cluster stars.  The remaining $\sim$10 lines were unmeasurable because of rotational broadening or because the line depths were comparable to the level of noise in the stellar continua.  Two Fe II lines could be measured for our stars, but the abundances derived from the lines proved to be unreliable in the cluster stars and in the Sun.  We therefore exclude Fe II from our analysis.  Other elements included in our analysis are Li I, Al I, Si I, Ca I, and Ni I.  The log gf values used for the handful of lines measured for these species were taken from Thevenin (1990), except for Li I. The Li I feature at $\lambda$6708 is a blend of Li features, both $^{6}$Li and $^{7}$Li isotopes, and an Fe I line.  The line list adopted in this region was kindly provided by C. Sneden (private communication), and contains Li wavelength and log gf values from Andersen et al. (1984), as originally published in Wiese et al. (1966).  Similarly, each line of the Al doublet at $\lambda$$\lambda$6696,6699 \AA\ is a blend of Al I and Fe I at our spectral resolution.

The equivalent width (EW) of each line was measured by fitting a Gaussian profile to the line profile with the IRAF task splot, and deblended from neighboring lines when necessary.  Depression of the continuum due to stellar rotation (in our earliest spectral types) and strong metal lines (in the later spectral types) is significant and must be considered when setting the local continuum level about each line.  The solar daytime sky spectrum observed with the same instrumental configuration was convolved with a Gaussian kernel so that the FWHM of solar lines was comparable to the FWHM of stellar lines of rotating stars.  Artificial noise characteristic of the T2KA detector was then added to match the S/N ratio of the observed spectrum.  For each spectral region, the continuum level of the noisy, artificially-broadened solar template spectrum was visually compared to that of the stellar spectrum for each measurement.  These comparisons were performed to avoid systematic offsets in continuum placement that might be introduced from stellar rotation, strong lines, and the limited S/N ratio of the stellar spectra.

Tables $\ref{tab6}$ and $\ref{tab7}$ list the atomic data and EW measurements for the stars analyzed in each cluster.  Repeated measurements of Fe I lines suggest that measurement uncertainties are usually less than 1 m\AA, and no more than $\sim$3 m\AA\ in the worst cases.  Another, more formal method to characterize the EW measurement uncertainty of a line, is given by Cayrel (1988), which includes consideration of the spectral resolution, the FWHM of the line, and the SNR of the observed spectrum.  Using this method, our EW uncertainties are $\sim$5.0 to 1.3 m\AA, for S/N ratios of $\sim$70--300.

\subsection{Atmospheric Parameters}

\begin{figure}
\epsscale{1.00}
\plotone{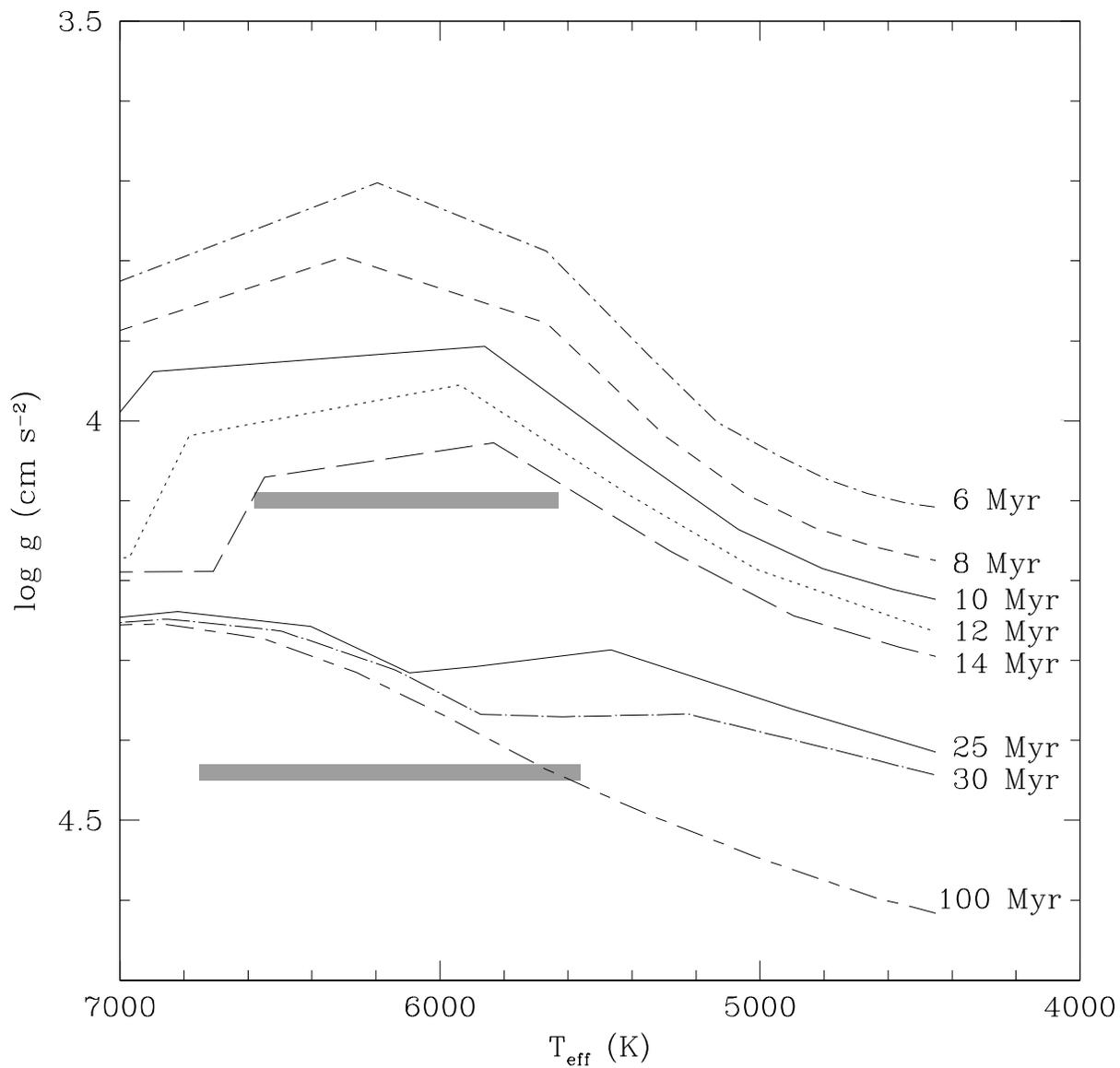}
\caption{A plot of log g vs. T$_{\rm eff}$ for pre-main-sequence tracks from Siess et al. (2000).  Values of log g were computed from the well-known equation, where values of L, M, and T$_{\rm eff}$ are provided from Siess et al.  The range of temperatures and adopted surface gravities for stars in NGC 7160 and NGC 2232 are indicated by the shaded regions.  Uncertainties in log g and T$_{\rm eff}$ are discussed in the text.}
\label{f6}
\end{figure}

Effective temperatures of program stars in NGC 7160 were determined from the spectral types provided by Sicilia-Aguilar et al. (2005) and from the temperatures and spectral types for dwarfs in Gray \& Corbally (2009).  Only four Fe II lines are available in our spectral region, of which only two or three lines were easily measured in the majority of our stars.  Iron abundances from Fe I and Fe II lines were determined for each star in 0.1 dex increments over 3.9 $\leq$ log g $\leq$ 4.5 dex, a suitable range of gravities for pre-main-sequence stars.  For the majority of stars, the best agreement of Fe I and Fe II abundances was found for log g = 4.1 dex.  A surface gravity of log g = 4.1 dex was adopted for all cluster stars to reduce scatter in our abundances, which could arise from difficulties in measuring weak, blended lines.  Pre-main-sequence tracks from Siess et al. (2000) over 6 to 14 Myr were used to confirm that log g = 4.1 dex is physically realistic for these young stars.  This range of isochrones was chosen because of the reported age spread of 4.6 Myr (s.d.) for NGC 7160 stars (Sicilia-Aguilar et al.).  A plot of log g versus T$_{\rm eff}$ was constructed from values of luminosity, mass, and T$_{\rm eff}$, as provided by Siess et al.; and is included in Figure $\ref{f6}$.  Our stars are typically $\sim$0.15--0.2 dex larger than the expected log g for 10 Myr old stars.  

Microturbulent velocities (v$_{t}$) were computed from the empirical relation for Galactic F and G dwarfs in Reddy et al. (2003), which depends on T$_{\rm eff}$ and log g.  Too few weak Fe I lines were available to determine v$_{t}$ from the spectra directly, at our spectral resolution.

Effective temperatures of stars in NGC 2232 were determined from UBV and JHK photometry, since no spectral types are available in the literature.  Temperatures were first calculated from dereddened (B-V) and (V-K) color indices using the color-temperature relations of Ramirez \& Melendez (2005).  Temperatures derived from the two color indices differed by $\sim$50--240 K for the stars.  

Since differences of $\sim$100 K can affect the derived metallicity by $\sim$0.08 dex (see \S 5.3), we constructed spectral energy distributions (SEDs) for the stars using the same photometric data.  The online SED fitting tool of Robitaille et al. (2007)\footnote{The SED fitting tool may be accessed at http://caravan.astro.wisc.edu/protostars/sedfitter.php} was used to fit model atmospheres to the photometry, and to derive temperatures for the stars.  The fitting tool allows a range of interstellar reddening to be considered for the best fits.  We used a value computed from the cluster average E(B-V) = 0.07 $\pm$ 0.02 (A$_{V}$ = 0.22), and a range bounded by the 1-$\sigma$ error of the average E(B-V) (0.16  $\leq$ A$_{V}$ $\leq$ 0.28), as determined by Lyra et al. (2006), and found that the assumed reddening did not change the best-fit temperatures. The best-fit model spectral energy distributions were inconsistent with the flux densities derived from published magnitudes in some filters by $\sim$15\%, which may explain the discrepancies obtained from the two color-temperature relations for some of the stars.  

Both methods of temperature determination (color-temperature and spectral energy distribution) are valid and we have no reason to favor either temperature scale, based on existing data for the stars.  Since abundances are sensitive to errors in temperature, we present two sets of abundances for NGC 2232 in Table \ref{tab9}; one computed from (V-K) and one obtained from the SED fitter.  The (V-K) photometric index was chosen over (B-V) because calibrations for this index are less sensitive to metallicity.  The average difference between the temperatures obtained from (V-K) and SED fitting is $\sim$160 K.  

A surface gravity of log g = 4.44 was used for all stars in NGC 2232, a value typical of main sequence stars.  The 25 and 100 Myr isochrones in Figure $\ref{f6}$ are close together for F-type stars, and differ by only $\sim$0.15--0.2 dex in log g for the G-type stars.  Our adopted surface gravities are within $\sim$0.15--0.2 dex of the theoretical values appropriate for young stars.  This level of uncertainty in the surface gravity does not significantly affect our abundance determinations, as we include only neutral species in our analysis.  Values of v$_{t}$ were computed using the same relation as for NGC 7160.

\subsection{Abundance Determinations}

Abundances for each cluster star are presented in Tables \ref{tab8} and \ref{tab9} for NGC 7160 and NGC 2232, respectively.  The elemental abundances for NGC 7160 were averaged over all lines for each species, relative to the solar abundance derived using the same line list.  The exception to this was Ca, for which we computed our abundances line-by-line differentially with respect to the Sun.  The reason for the exception was the large line-to-line scatter ($\sim$0.3 dex) found for both the Sun and our cluster stars, likely due to the log gf values used.  Also included in Tables \ref{tab8} and \ref{tab9} are the standard deviations and number of lines used for each element.  Abundances of NGC 2232 were determined in the same manner as for NGC 7160, except that Ni abundances were also computed line-by-line differentially with respect to the Sun.  The ``blends" driver of MOOG was employed for the blended lines of Li I and the Al I doublet.  The [el/Fe] ratios are computed using the corresponding [Fe/H] value for each star.

In Table \ref{tab10} we present typical abundance uncertainties due to errors in atmospheric parameters.  The last column in the table includes a total uncertainty for each element, obtained by adding in quadrature the uncertainties of T$_{\rm eff}$, log g, and v$_{t}$.  The cumulation of uncertainties for these parameters results in an uncertainty of $\sim$0.1 dex in [Fe/H]. The most significant contributors to the uncertainties in our results are T$_{\rm eff}$ and v$_{t}$, which contribute comparable amounts to the overall uncertainty, over the parameter space we spanned.

\section{Results}

\subsection{Iron}

\begin{figure}
\epsscale{1.00}
\plotone{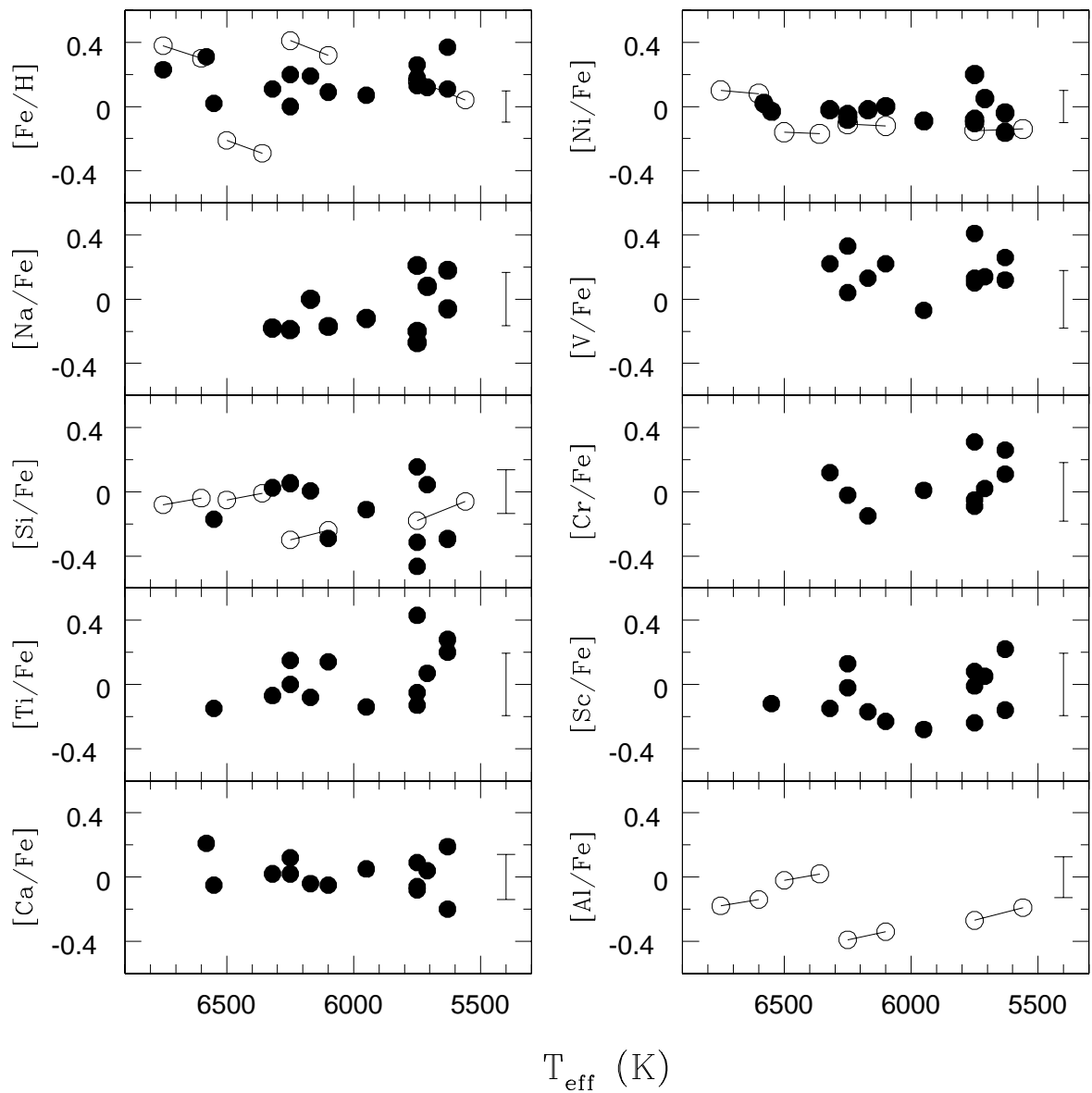}
\caption{[Fe/H] and [el/Fe] ratios for stars in NGC 7160 (solid points) and NGC 2232 (open points). The two sets of abundances derived for NGC 2232, using the two temperature scales, are joined by a solid line.  Representative error bars are included on the right of each panel.}
\label{f7}
\end{figure}

NGC 7160 and NGC 2232 are both found to be super-solar in [Fe/H].  Figure \ref{f7} presents elemental abundances as a function of T$_{\rm eff}$ for stars in both clusters.  A representative error bar for each element is included on the right of each panel, which includes both the standard deviation of each element (s.e.m. for species with more than five lines) and uncertainties due to model atmosphere parameters, added in quadrature.  The two sets of abundances (determined for the two temperature scales) for each star of NGC 2232 are shown connected by solid lines.  

The [Fe/H] abundances of NGC 7160 stars are all above solar with a cluster average of 0.16 $\pm$ 0.03 (s.e.m.).  The high Fe abundance of this cluster places it in a metallicity regime along with several other young clusters (e.g., NGC 6475, [Fe/H] = 0.14, Sestito et al. 2003; Hyades, [Fe/H] = 0.13, Paulson et al. 2003; and NGC 6134, [Fe/H] = 0.15, Carretta et al. 2004).  The range in metallicity for stars presumed to be members of this cluster can be explained by observational uncertainties.  Given the difficulty of establishing membership  via radial velocities, we excluded stars more than 1-$\sigma$ from the median value of the stars analyzed, and recomputed the cluster average to gauge the possible impact that non-membership may have on our average.  The resulting value is  0.15 $\pm$ 0.04 (s.e.m.), which suggests that the inclusion of possible non-members is not strongly biasing our result.  Another possible concern is the inclusion of three WTTSs (01-615, 03-654, and 03-835) in the analysis, because of possible veiling effects due to accretion that are present in the spectra of classical T-Tauri stars.  The three WTTSs have high metallicities ([Fe/H] = 0.18--0.37) on the cool end of the temperature distribution, but are within the range of abundances for the other stars in our sample.  While veiling is not expected in our WTTSs, our results appear to confirm that veiling is not a concern, since veiling would fill in the spectral lines and produce abundances lower than the cluster average.

NGC 2232 is also found to be super-solar, with a cluster average of [Fe/H] = 0.09 $\pm$ 0.14 (s.e.m.) or 0.18 $\pm$ 0.14 (s.e.m.) for four stars, depending on the temperature scale used.  Inspection of [Fe/H] abundances in NGC 2232 reveals a spread in metallicity, with two stars at [Fe/H] = 0.3--0.4 (18588 and 18569), one star at [Fe/H] = 0.04--0.16 (18585), and another at [Fe/H] = -0.21 to -0.29 (9242).

Recall that the presence of the Li $\lambda$6708 \AA\ absorption line was used as a membership criterion for this cluster, due to the range in the radial velocities of stars in the field of NGC 2232.  Inspection of the radial velocities of the four stars analyzed shows that the three stars with super-solar abundances have velocities $\sim$26 km s$^{-1}$, whereas, the sub-solar star 9242 has a V$_{r}$$\sim$53 km s$^{-1}$.   Our metallicity results may suggest that there is a true cluster mean at 26.6 $\pm$ 0.77 km s$^{-1}$, and that star 9242 is possibly a nonmember.  The presence of young foreground or background sources is not unreasonable in this field, given the number of X-ray detections reported that are not consistent with a 25 Myr isochrone (Currie et al. 2008b).  If we remove star 9242 from our calculation of the cluster mean, the average cluster abundance increases to [Fe/H] = 0.22 $\pm$ 0.09 or 0.32 $\pm$ 0.08.  Abundances reported for NGC 2232 in subsequent sections are averages for the three stars with V${_r}$$\sim$26 km s$^{-1}$, and exclude star 9242.  Mean cluster abundances are presented for both NGC 7160 and NGC 2232 in Table \ref{tab11}.

\subsection{Lithium}
Lithium abundances for NGC 2232 are presented in Table \ref{tab9}.  The abundances of the two most metal-rich stars are in agreement, both with log $\epsilon$ (Li) = 2.99 or 3.10, depending on the temperature scale used.  These values are also in agreement with Li abundances reported for stars with similiar T$_{\rm eff}$ in other young clusters, whose average is log $\epsilon$ (Li) $\simeq$ 3.1--3.3 (Travaglio et al. 2001 and references therein).  Abundances of these young clusters are assumed to trace the present Li ISM abundances, since the stars have undergone little, if any, pre-main-sequence or main sequence Li depletion (Pinsonneault 1997).  

\begin{figure}
\epsscale{1.00}
\plotone{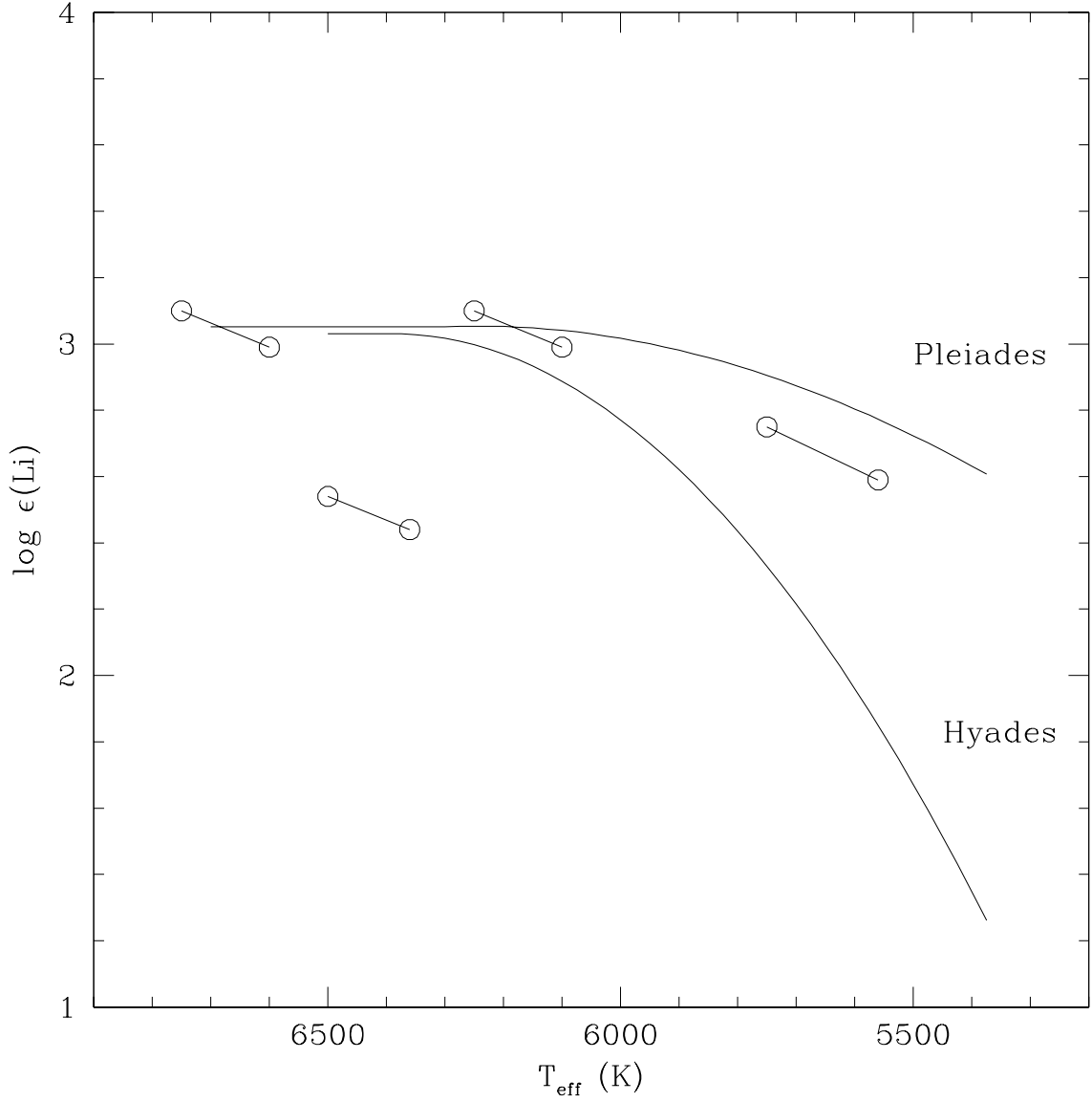}
\caption{Lithium abundances of stars in NGC 2232, with curves representing the average Li abundances of stars in the Pleiades (King et al. 2000) and Hyades (Soderblom et al. 1993).}
\label{f8}
\end{figure}

In Figure \ref{f8} we plot Li abundance versus T$_{\rm eff}$ of the four stars, along with average Li distributions for two well-studied clusters, the Pleiades (120 Myr; King et al. 2000) and the Hyades (600 Myr; Soderblom et al. 1993).  The Li abundances of stars hotter than 6000 K are consistent with both the Pleiades and the Hyades, with the exception of star 9242.  The low Li abundance of this star adds further doubt that it is a member of NGC 2232, since no cluster younger than $\sim$200 Myr has been found to have the Li-dip reported for older clusters like the Hyades (Soderblom et al. 1993).  The position of the coolest star in our sample, 18585, is lower than the average Li abundance of stars in the Pleiades at the same temperature.  More depletion of Li in star 18585 might be expected even at an age of only 25 Myr, since higher abundances increase the opacity of the gas and cause a deepening of the stellar convection zone, where the hotter temperatures burn Li more readily (Pinsonneault 1997 and references therein).  The amount of surface Li depletion expected for a T$_{\rm eff}$ $\sim$5560--5750 K star, or $\sim$1.1--1.2 M$_{\odot}$ star (according to the 25 Myr Siess et al. 2000 isochrone), at 25 Myr is $\sim$0.20--0.35 dex (Pinsonneault et al. 1990).  Li isochrones at 100 Myr suggest a similar level of depletion, $\sim$0.2--0.3 dex, for an increase in [Fe/H] from 0.00 to 0.15 (Pinsonneault 1997).

\subsection{$\alpha$-elements: Si, Ca, and Ti}
In Figure \ref{f7} we include abundances of available $\alpha$-elements in our spectral regions.  The [Si/Fe] abundances of NGC 7160 are averages of the [Si I/Fe] and [Si II/Fe] abundances for each star.  The average [Si/Fe] abundance, computed for all Si measurements, is slightly sub-solar, at -0.12 $\pm$ 0.04 (s.e.m.).  Averages for the individual species of Si are offset by $\sim$0.1 dex when compared to Fe I, with  [Si I/Fe] = -0.06 $\pm$ 0.04 and [Si II/Fe] = -0.19 $\pm$ 0.07.  The Si I and Si II abundances are in closer agreement when Fe II is used to compute the [Si II/Fe] ratio, as the average ratio increases to [Si II/Fe II] = -0.08 $\pm$ 0.04.  The remaining two $\alpha$-elements in NGC 7160 have solar abundances:  [Ca/Fe] = 0.02 $\pm$ 0.03 (s.e.m.) and [Ti/Fe] = 0.05 $\pm$ 0.05.

For NGC 2232 the available $\alpha$-elements included Si I and Ca I.  The average Si abundances for the three stars thought to be members with the two sets of temperatures are -0.11 $\pm$ 0.06 and -0.19 $\pm$ 0.07, results that put the [$\alpha$/Fe] ratio at a slightly sub-solar level.  Unfortunately, the available Ca I lines in our spectral region are too strong for a reliable analysis.

\subsection {Na and Al}
NGC 7160 was found to have a slightly sub-solar abundance of the light element Na, with a mean of [Na/Fe] = -0.07 $\pm$ 0.05 (s.e.m.).  Our value is solar within 1-$\sigma$ of the mean, and is consistent with the roughly solar [Na/Fe] abundances from other studies of dwarfs in open clusters (e.g., Hyades, Paulson et al. 2003; IC 4651, Pasquini et al. 2004).  The solar abundances of Na are in contrast to the $\sim$0.1--0.2 dex enhancements reported for many studies of giants in open clusters in the galactic disk (e.g., Jacobson et al. 2009).  The differences between derived [Na/Fe] distributions of dwarfs and giants are part of an unresolved problem that was first discussed by Cayrel de Strobel et al. (1970).  Sodium abundances of solar-metallicity dwarfs derived from the $\lambda$$\lambda$6154, 6160 doublet are not expected to be subject to substantial NLTE effects (Gratton et al. 1999; Mashonkina et al. 2000); therefore, we do not apply any corrections to the results presented in Table \ref{tab8}.

Aluminum abundances were determined for stars in NGC 2232.  The average abundances for the two sets of temperatures are [Al/Fe] = -0.22  $\pm$ 0.06 (s.e.m.) and -0.28 $\pm$ 0.06.  Two of the stars are roughly solar within one standard deviation of the mean (0.1 dex).  Abundances from the non-resonance  $\lambda$$\lambda$6696, 6698 doublet lines are not expected to suffer from appreciable NLTE effects for the spectral and luminosity range of stars in our sample (Baumueller \& Gehren 1997), so we do not apply any corrections. 

\begin{figure}
\epsscale{1.00}
\plotone{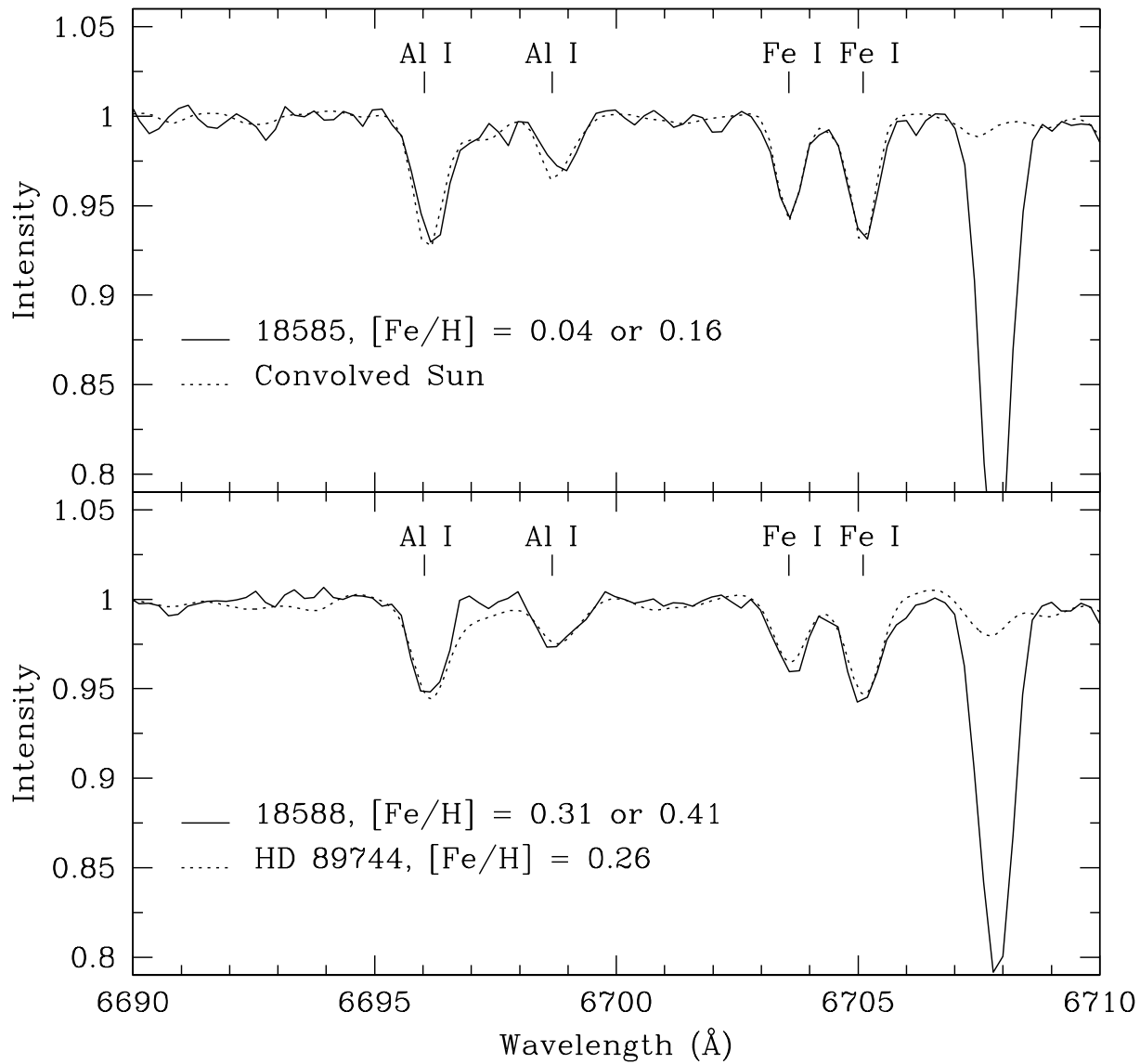}
\caption{Spectra of the $\lambda$$\lambda$6696,6698 Al doublet in NGC 2232 stars 18585 and 18588.  The cluster stars are compared to spectra of stars with similar metallicities and T$_{\rm eff}$.  Top panel:   Spectrum of star 18585 and a convolved solar spectrum. Bottom panel:  Spectrum of star 18588 and a convolved spectrum of HD 89744 obtained from the ELODIE spectral database.}
\label{f9}
\end{figure}

In Figure \ref{f9} we compare the $\lambda$$\lambda$6696, 6698 Al doublet lines of stars 18585 ([Fe/H]  = 0.04 or 0.16) and 18588 ([Fe/H] = 0.31 or 0.41) to stars with similar metallicities and T$_{\rm eff}$, the Sun and HD 89744 ([Fe/H] = 0.26, T$_{\rm eff}$= 6291 K, log g = 4.07; Valenti \& Fischer 2005), respectively.  The blended Al lines of our cluster stars both visually match their comparison stars well.  The spectrum of HD 89744 in Figure \ref{f9} was obtained from the ELODIE spectral archive\footnote{See http://atlas.obs-hp.fr/elodie/.}, and has been broadened to match our spectral resolution.  However, we found no [Al/Fe] abundance determined from the $\lambda$$\lambda$6696, 6698 doublet of the ELODIE spectrum in the literature.  Reported ratios for HD 89744 include [Al/Fe] = 0.08 using $\lambda$$\lambda$8772,8773 (for [Fe/H] = 0.18; Edvardsson et al. 1993), [Al/Fe] = -0.12 using $\lambda$$\lambda$7835,7836 (for [Fe/H] = 0.30; Gonzalez et al. 2001), and [Al/Fe] = 0.14 using an average of the $\lambda$$\lambda$6696, 6698; $\lambda$$\lambda$7835,7836; and $\lambda$$\lambda$8772,8773 doublets (for [Fe/H] = 0.22; Beirao et al. 2005).   If the Gonzalez et al. [Al/Fe] ratio is correct, then star 18588 may actually be slightly sub-solar.  However, the slightly positive [Al/Fe] ratios found by Beirao et al. and Edvardsson et al. may suggest our derived [Al/Fe] ratio for 18588 is too low.  The convolved solar spectrum visually matches the spectrum of star 18585 well.  Our derived solar Al abundance from our daytime sky spectrum is 0.11 higher than the standard value reported by Anders \& Grevesse (1989).

In an effort to further understand why our [Al/Fe] ratios are sub-solar, we compared the Al I lines of  synthetic spectra of 6100 K and 6250 K stars with log g = 4.44 and [Fe/H] = 0.3.  For a fixed [Al/Fe] = -0.34, we found that the EWs of the blended doublet lines only changed by $\sim$10\% ($\sim$3--4 m\AA), suggesting that our temperature uncertainty is not the reason for our low [Al/Fe] ratios.  To understand the effects of the input metallicity of our model atmosphere on our [Al/Fe] abundances, we compared synthetic spectra for models at [Fe/H] = 0.0 and 0.3 at 6100 K, for a fixed log $\epsilon$ (Al) value.  The EWs of the doublet lines, which are each blends of Al I and Fe I, changed by $\sim$20\% ($\sim$5--9 m\AA), suggesting that our Al abundances are more sensitive to uncertainties in Fe than temperature.

\subsection{Fe-peak elements:  Ni, V, Cr, and Sc}
The Fe-peak elements Ni, V, Cr, and Sc were included in our analysis of NGC 7160.  As expected Ni, Cr, and Sc scale with Fe, with abundances of [Ni/Fe] = -0.03 $\pm$ 0.02, [Cr/Fe] = 0.05 $\pm$ 0.05, and [Sc II/Fe] = -0.07 $\pm$ 0.04.  The Ni abundances were determined from three lines for the earliest types and $\sim$5--11 lines for the later spectral types, and follow a solar distribution.  The typical line-to-line scatter is $\sim$0.2 dex for the latest spectral types.  The Cr abundances were obtained from only one measured line, and are therefore uncertain.  Similarly, the mean Sc II abundance was obtained from one line per star.  Vanadium abundances were obtained from one to three lines, with a cluster average of [V/Fe] = 0.17 $\pm$ 0.04 (s.e.m.).  The fact that [V/Fe] is relatively constant for stars in our sample, despite the high temperature sensitivity of V lines, suggests that our adopted temperatures are reliable.

The Ni abundances of NGC 2232 stars are [Ni/Fe] = -0.06 $\pm$ 0.07 and -0.06 $\pm$ 0.08 for the two temperature scales, which are solar within the errors.  These results are fairly robust, as they were determined from three to five lines per star; however, the scatter in the individual line abundances is $\sim$0.2 dex per star.  

\section{Discussion}

\subsection{Comparison to Previous Studies}

Daflon et al. (1999) carried out a LTE iron abundance analysis of slowly-rotating O and B stars in the Cepheus OB2 region.  The subset of stars for which they were able to derive abundances included five stars from Tr 37 (the younger cluster in Cep OB2), and three stars ``in the general region of the older subgroup of Cep OB2."  The positional offsets of the three stars in or near NGC 7160 (HD 206327, HD 239742, and HD 239743) from the center of NGC 7160 are between 2.0$\degr$--5.7$\degr$, making them not likely members of NGC 7160, which has a diameter of $\sim$0.5$\degr$ (from bright, early-type members; Kharchenko et al. 2005) at a distance of 900 pc (Sicilia-Aguilar et al. 2005), but possibly members of Tr 37.  The size of NGC 7160 is probably an underestimate, as the entire region around the cluster has not been deeply imaged, and Sicilia-Aguilar et al. assert that they may not have fully mapped the main area of the cluster, and low-mass stars in the cluster may be located further from the center.  

In the interest of learning more about the chemical composition of stars in Cep OB2, we consider the three stars near NGC 7160 by Daflon et al. (1999).  They derive sub-solar metallicities from five Fe III lines:  [Fe/H] = -0.28\footnote{Assuming log $\epsilon$ (Fe)$_{\odot}$ = 7.52.} $\pm$ 0.10, -0.39 $\pm$ 0.14, and -0.05 $\pm$ 0.11 for HD 206327, HD 239742, and HD 239743, respectively.  The difference in average values of [Fe/H] between our study and that of Daflon et al. is large, at 0.4 dex.  The reason for this discrepancy remains unclear.  A resolution of the discrepancy will require further observational work on a larger sample of stars known to be members.  Determinations of nebular abundances for the Cep OB2 region would also be useful.

The only study of NGC 2232's metallicity available in the literature was presented by Pastoriza \& Ropke (1983), using intermediate-band colors to define a CN index.  These authors constrained the metallicity to [Fe/H] = -0.20 from two giant stars thought to be members of the cluster at that time.  Claria et al. (1985) showed that the giants did not satisfy criteria for membership in the young cluster, however, based on the position of the stars ``below the main sequence termination point," and one giant having a  radial velocity offset $\sim$60 km s$^{-1}$ from velocities of early-type members.

\subsection{Comparison to Other Metal-Rich Stars}

\begin{figure}
\epsscale{1.00}
\plotone{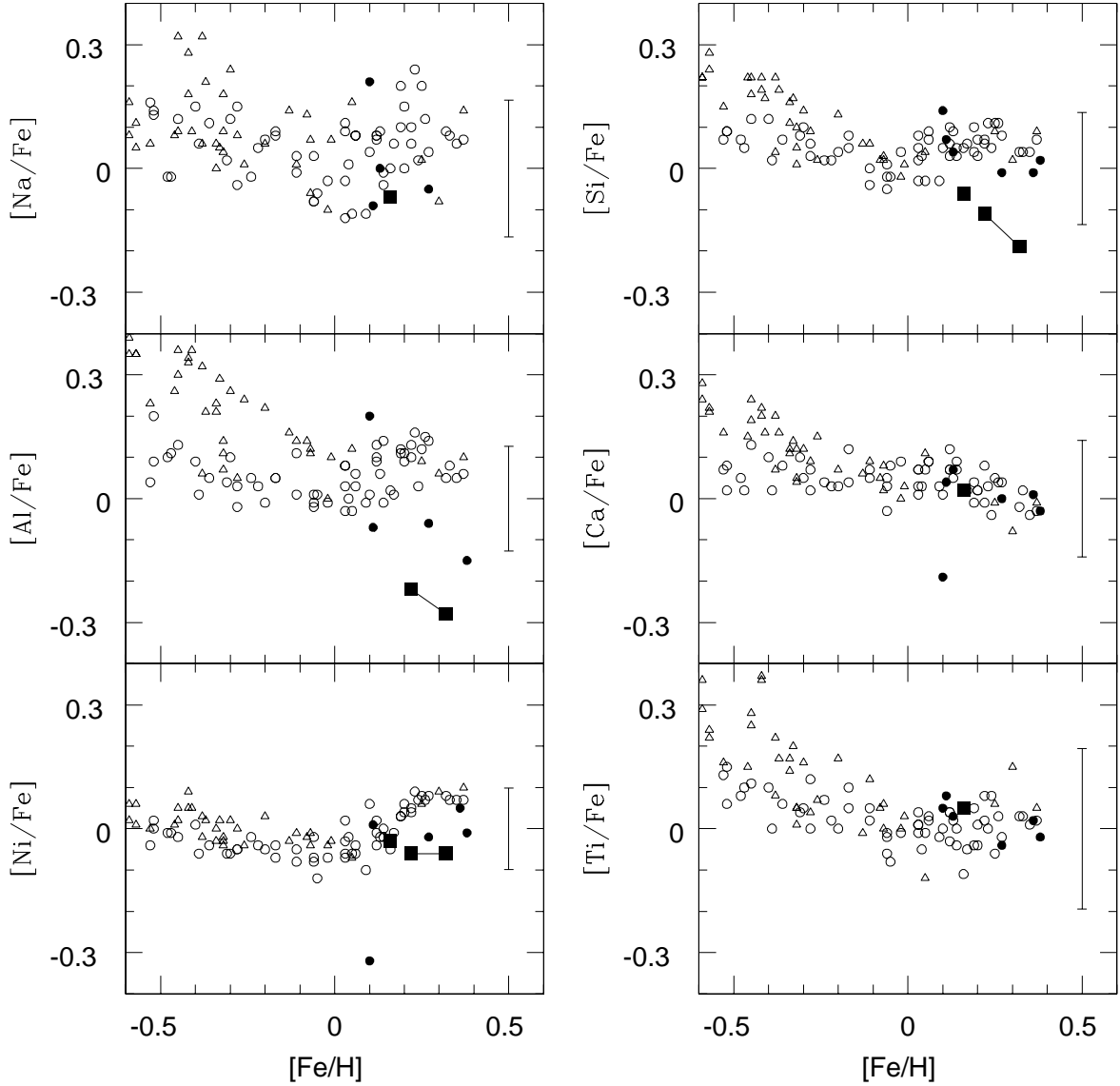}
\caption{Elemental abundances relative to Fe of Galactic thin and thick disk stars and other metal-rich open clusters.  NGC 7160 and NGC 2232 are represented by filled squares, with the two abundances of NGC 2232 joined by a solid line.  Representative error bars are included on the right of each panel.  Galactic thin disk stars and thick disk stars from Bensby et al. (2005) are represented by open circles and triangles, respectively.  Transitional stars between the thin and thick disk are also represented by triangles.  Open clusters are represented by filled circles.  Open cluster data were obtained from many sources, described in the text.}
\label{f10}
\end{figure}

In Figure \ref{f10} we compare the elemental abundances of NGC 7160 and NGC 2232 with Galactic thin and thick disk stars and other metal-rich open clusters.  The thin and thick disk stars were obtained from the full sample of Bensby et al. (2005).  Abundances were compiled for the following open clusters with [Fe/H] $>$ 0.1:  IC 4651 dwarfs ([Fe/H] = 0.11, 1.7 Gyr; Pasquini et al. 2004), Praesepe ([Fe/H] = 0.27, 600 Myr; Pace et al. 2008), NGC 6705 ([Fe/H] = 0.10, 150 Myr; Gonzalez \& Wallerstein 2000), NGC 6791 ([Fe/H] = 0.38, 8 Gyr; Carraro et al. 2006), NGC 6253 ([Fe/H] = 0.36, 3 Gyr; Sestito et al. 2007), and the Hyades ([Fe/H] = 0.13, 600 Myr; Paulson et al. 2003).  No effort was made to standardize the abundances to the same solar value of iron.

The average [Na/Fe] abundance of NGC 7160 was found to be slightly sub-solar, a result that is consistent with three of the four other metal-rich open clusters.  The range of [Na/Fe] $\sim$-0.1 to 0.0 of these clusters, is slightly lower than the average Na abundance of disk stars at the same metallicities (NGC 6705 deviates from this average).  The average disk Na abundance increases for [Fe/H] $>$ 0, and has an average [Na/Fe]$\sim$0.1, with several tenths of a dex dispersion.

Interestingly, the [Al/Fe] ratios for three of the literature open clusters are low compared to the disk stars.  The metal-rich open clusters have a reported range of [Al/Fe] from $\sim$ -0.07 to -0.15 dex.  NGC 2232 is $\sim$0.1--0.15 dex below the other clusters, a difference that is perhaps systematic, but is also within the standard deviation of Al abundances for the cluster.  One result that stands out in Figure \ref{f10} is that this small set of metal-rich clusters appears to be underabundant in Al compared to the sample of disk stars, which show [Al/Fe] $\sim$ 0.1--0.2 for [Fe/H] $>$ 0.  Whether this apparent deficiency of Al is real or results from overcorrection of the blended Fe I lines remains unclear.

Both NGC 7160 and NGC 2232 are determined to have roughly solar [Ni/Fe] abundances within our reported errors.  The other open clusters also have solar ratios (again, excluding NGC 6705), a result that appears to be slightly discrepant from the field stars at the same metallicities.  The disk stars have a slight, $\sim$0.1 dex overabundance for [Fe/H] $>$0.1. 

The $\alpha$-element abundance ratios of [Si/Fe], [Ca/Fe], and [Ti/Fe] in NGC 7160 are solar within the errors, and are indistinguishable from the other open clusters and disk star populations.  Here we compare the [Si/Fe] abundance from only Si I lines for NGC 7160 to the other samples, for consistency.  The Si abundance of NGC 2232 is underabundant compared to the other clusters and the disk stars.  This difference may result from systematic errors, but it is interesting to note the similarities between the Si and Al distributions for the disk stars and NGC 2232.  As noted by McWilliam (1997) and Bensby et al. (2003), the distribution of [Al/Fe] abundances for disk stars follows the same trend exhibited by $\alpha$-elements, across the full metallicity range of their studies.  More specifically, the disk stars show declines in [Si/Fe] and [Al/Fe] excesses as metallicity approaches [Fe/H] $\sim$ 0, and then a slight increase for Si and Al at [Fe/H] $>$ 0.  NGC 2232 is underabundant in both species, and thus qualitatively exhibits the same correlation of Al and Si.

In summary, abundances of NGC 7160 are consistent with other metal-rich open clusters and Galactic thin and thick disk stars, within in our observational errors.  The Al and Si abundances of NGC 2232 are underabundant compared to the sample of disk stars.  Whether our [Si/Fe] abundances are lower than the other samples because of systematic effects remains to be seen.    Our results are obtained from pre-main-sequence stars, while the metal-rich cluster abundances were derived from stars at a variety of evolutionary states, including dwarfs (IC 4651, Hyades, and Praesepe), giants (NGC 6791 and NGC 6253), and supergiants (NGC 6705) (see references above).  Model atmosphere deficiencies and departures from LTE may affect dwarfs and giants differently, and thus mask any actual differences or trends for the observed clusters.

\subsection{Spitzer Observations of Target Stars}

Two of the F stars in NGC 2232 for which we present abundances, were detected at 24 $\micron$.  Stars 18569 and 18588, the two stars with the highest metallicities, have K$_{S}$ - [24] indices of 0.28 and 0.59, respectively (Currie et al. 2008b).  According to the criteria used by Currie et al. to identify stars with robust 24 $\micron$ excesses (the excess must be greater than three times the [24] error and comprise more than 15\% of the observed flux) and their reported photometric error, star 18588 would be an excess star and a debris disk candidate.  However, star 18588 is not included in the final candidate membership list of Currie et al., who used V vs. (V-J) and 2MASS NIR CMDs to select members for their Spitzer analysis.  Star 18588 is an X-ray active star that was included in their initial J vs. (J-K$_{S}$) diagram, but it was slightly too faint in the J-band or slightly too blue in NIR colors to be included as a candidate member in their study.  It is interesting to note that star 18588 has the lowest Si and Al abundances in our small sample of stars in NGC 2232.  

None of our chosen targets in NGC 7160 were detected in Spitzer 24 $\micron$ observations (Sicilia-Aguilar et al. 2006).

\section{Summary \& Conclusions}

We have determined radial velocities and abundances of Fe-peak, $\alpha$-elements, and light elements for F- and G-type stars in the young open clusters NGC 7160 and NGC 2232, using moderate resolution spectra (R = 16,000) obtained with the multi-object spectrograph Hydra on the WIYN 3.5-m telescope.  Our analysis shows that both clusters have super-solar metallicities, with [Fe/H] = 0.16 $\pm$ 0.03 (s.e.m.) and  0.22 $\pm$  0.09 (s.e.m.) or 0.32 $\pm$  0.08 (s.e.m.) for NGC 7160 and NGC 2232, respectively.

The main results of our analysis are summarized below.

--Radial velocities of stars in NGC 7160 do not appear to have a clear peak in the distribution, and exhibit a large breadth in velocity, suggesting that the cluster is not gravitationally bound.

--The average radial velocity of NGC 2232 was found to be 26.6 $\pm$ 0.77 km s$^{-1}$, based on velocities of four stars with Li absorption lines.

--The [$\alpha$/Fe] ratios of stars in NGC 7160 are roughly solar and are indistinguishable from other metal-rich open clusters and disk stars with similar metallicities.  Stars in NGC 7160 exhibit [Na/Fe] and [Ni/Fe] abundances that are slightly lower than disk stars, but are in agreement with other metal-rich clusters.

--The [Al/Fe] and [Si/Fe] ratios of NGC 2232 stars may be underabundant compared to thin and thick disk stars with the same metallicities.  Other metal-rich open clusters have similarly low Al abundances, as well.  The [Ni/Fe] ratios for this cluster are solar, and as for NGC 7160, are consistent with other metal-rich clusters and are slightly lower than thin and thick disk stars.

\acknowledgments

Support for this work was provided by NASA through a contract issued by the Jet Propulsion Laboratory, California Institute of Technology under a contract with NASA.  Funding for this work was also provided through a NASA Graduate Student Researchers Program (GSRP) Fellowship from Goddard Space Flight Center.  We would like to thank George Jacoby for providing director's discretionary time to obtain a portion of these observations on the WIYN 3.5-m telescope.  Pilachowski gratefully acknowledges support of the College of Arts and Science at Indiana University through the Daniel Kirkwood Chair.  We would like to thank the anonymous referee for a careful reading of this manuscript and for suggested improvements to the text, as well as Luisa Rebull and Christian I. Johnson for useful discussions.  This publication makes use of data products from the Two Micron All Sky Survey, which is a joint project of the University of Massachusetts and the Infrared Processing and Analysis Center/California Institute of Technology, funded by the National Aeronautics and Space Administration and the National Science Foundation.  This research has also made use of the WEBDA database, operated at the Institute for Astronomy of the University of Vienna.  Lastly, part of this work is based on spectral data retrieved from the ELODIE archive at Observatoire de Haute-Provence (OHP).

{\it Facilities:}\facility{WIYN (HYDRA)}.

\clearpage

\clearpage

\clearpage

\clearpage

\tablecolumns{4}
\tablewidth{0pt}

\begin{deluxetable}{lcll}
\tabletypesize{\footnotesize}
\tablecaption{Hydra Observations of Target Clusters \label{tab1}}

\tablehead{
\colhead{Cluster}      &
\colhead{Wavelength ($\AA$)}      &
\colhead{UT Date}      &
\colhead{Exposure (s)}      

}

\startdata

NGC 7160  & 6280  & 2006 Dec 04 & 1 x 3600  \\
  & 6280  & 2006 Dec 05 & 2 x 3600  \\
  & 6280  & 2006 Dec 06 & 4 x 3600  \\
NGC 2232  & 6600  & 2007 Feb 27 & 2 x 2700  \\
  & 6600  & 2007 Feb 27 & 3 x 3600  \\
  & 6600  & 2007 Mar 01 & 1 x 3600  \\
  & 6600  & 2007 Mar 01 & 1 x 2100  \\

\enddata


\end{deluxetable}

\clearpage
\tablecolumns{2}
\tablewidth{0pt}

\begin{deluxetable}{ll}
\tabletypesize{\footnotesize}
\tablecaption{Continuum Regions in the Observed Spectra \label{tab2}}

\tablehead{

\colhead{NGC 7160}  &
\colhead{NGC 2232}  \\

\colhead{Wavelength ($\AA$)}      &
\colhead{Wavelength ($\AA$)}           

}

\startdata

6139.1--6140.6  & 6386.7--6391.4  \\
6142.6--6144.0  & 6401.7--6404.8  \\
6152.5--6153.6  & 6441.8--6447.3  \\
6196.2--6198.2  & 6505.0--6507.3  \\
6222.1--6223.3  & 6538.9--6541.6  \\
6234.1--6235.9  & 6588.3--6590.7  \\
6386.2--6387.9  & 6620.3--6623.1  \\
6388.8--6389.9  & 6648.7--6652.8  \\
6427.1--6429.4  & 6673.1--6676.4  \\
\nodata & 6681.9--6686.9  \\

\enddata

\end{deluxetable}

\clearpage
\tablecolumns{7}
\tablewidth{0pt}

\begin{deluxetable}{llllccl}
\tabletypesize{\footnotesize}
\tablecaption{Radial Velocities of Stars in NGC 7160 Field \label{tab3}}

\tablehead{
\colhead{Star}      &
\colhead{SpT}      &
\colhead{V}      &
\colhead{B-V}      &
\colhead{Mem}      &
\colhead{V$_{r}$ (km s$^{-1}$)}      &
\colhead{$\sigma$$_{r}$ (km s$^{-1}$)}
        
}

\startdata

DG-40 & F7.0  & 12.59 & 0.78  & Y & 4.3 & 0.3 \\
DG-41 & F2.0  & 12.64 & 0.68  & Y & -39.1 & 0.5 \\
DG-55 & F3.0  & 13.39 & 0.81  & Y & -15.2 & 0.9 \\

\enddata

\tablecomments{Star IDs, spectral types, membership determinations, and photometry are from Sicilia-Aguilar et al. (2005).   Stars considered members, probable members, and probable nonmembers are represented by ``Y," ``P," and ``PN," respectively.  See text.  A full version of the data table can be made available upon email request.}

\end{deluxetable}

\clearpage
\tablecolumns{8}
\tablewidth{0pt}

\begin{deluxetable}{llllcccl}
\tabletypesize{\footnotesize}
\tablecaption{Radial Velocities of Stars in NGC 2232 Field \label{tab4}}

\tablehead{
\colhead{Star}      &
\colhead{V}      &
\colhead{B-V}      &
\colhead{Li}      &
\colhead{Mem}      &
\colhead{V$_{r}$ (km s$^{-1}$)}      &
\colhead{$\sigma$$_{r}$ (km s$^{-1}$)}         &
\colhead{Notes}  

}

\startdata

18569 & 11.47 & 0.53  & Y & Y & 26.7  & 0.9 & X-ray active,  Cand. from Lit.  \\
18585 & 12.49 & 0.71  & Y & Y & 26.3  & 0.4 & X-ray active,  Cand. from Lit.  \\
18588 & 12.13 & 0.62  & Y & Y & 25.9  & 1.0 & X-ray active  \\

\enddata

\tablecomments{Star IDs and X-ray detections are from Currie et al. (2008b).  Photometry is from Lyra et al. (2006).  Stars considered candidate members by Currie et al. are indicated by ``Cand. from Lit."  A full version of the data table can be made available upon email request.}

\end{deluxetable}

\clearpage
\tablecolumns{8}
\tablewidth{0pt}

\begin{deluxetable}{lccccccc}
\tabletypesize{\footnotesize}
\tablecaption{Radial Velocities of Binaries in NGC 2232 Field \label{tab5}}

\tablehead{
\colhead{Star}      &
\multicolumn{7}{c}{V$_{r}$ (km s$^{-1}$) for HJD(-2450000)}  \\

\hline

\colhead{}      &
\colhead{4158.63549}      &
\colhead{4158.67024}      &
\colhead{4158.70952}      &
\colhead{4158.75434}      &
\colhead{4158.79782}      &
\colhead{4160.69482}      &
\colhead{4160.72998}


}

\startdata
11242A  & 11.1  & 11.4  & 11.4  & 11.4  & 12.0  & 15.2  & 15.6  \\
11242B  & 58.5  & 58.0  & 57.6  & 57.3  & 56.5  & 54.2  & 52.8  \\
18650A  & 66.9  & 63.6  & 64.3  & 65.2  & 63.2  & 58.4  & 57.4  \\
18650B  & -9.6  & -11.3 & -12.2 & -12.8 & -12.3 & -11.4 & -11.7 \\

\enddata

\end{deluxetable}

\clearpage
\setlength{\voffset}{1.8cm}
\setlength{\tabcolsep}{0.05in} 
\tablecolumns{20}
\tablewidth{0pt}

\begin{deluxetable}{llllllllllllllllllll}
\rotate
\tabletypesize{\scriptsize}
\tablecaption{Atomic Line Data and Equivalent Widths for NGC 7160\label{tab6}}

\tablehead{
\colhead{Element}      &
\colhead{$\lambda$}      &
\colhead{LEP}      &
\colhead{log gf}      &
\colhead{Ref}      &
\colhead{DG-40}      &
\colhead{DG-55}      &
\colhead{DG-61}      &
\colhead{DG-62}      &
\colhead{DG-64}      &
\colhead{DG-349}      &
\colhead{DG-371}      &
\colhead{DG-423}      &
\colhead{DG-455}      &
\colhead{DG-644}      &
\colhead{DG-895}      &
\colhead{DG-921}      &
\colhead{01-615}      &
\colhead{03-654}      &
\colhead{03-835}       \\

\colhead{}      &
\colhead{($\AA$)}      &
\colhead{(eV)}      &
\colhead{}      &
\colhead{}  &
\colhead{ (m$\AA$) } &
\colhead{ (m$\AA$) } &
\colhead{ (m$\AA$) } &
\colhead{ (m$\AA$) } &
\colhead{ (m$\AA$) } &
\colhead{ (m$\AA$) } &
\colhead{ (m$\AA$) } &
\colhead{ (m$\AA$) } &
\colhead{ (m$\AA$) } &
\colhead{ (m$\AA$) } &
\colhead{ (m$\AA$) } &
\colhead{ (m$\AA$) } &
\colhead{ (m$\AA$) } &
\colhead{ (m$\AA$) } &
\colhead{ (m$\AA$) }

}

\startdata

Na I & 6154.23 & 2.10 & -1.66 & 1 & 17 & \nodata & 46 & \nodata & \nodata & 26 & 62 & \nodata & 71 & 29 & 59 & 21 & 37 & 41 & 66 \\
 Na I & 6160.75 & 2.10 & -1.35 & 1 & \nodata & \nodata & 49 & 18 & \nodata & 53 & 91 & 24 & 80 & 36 & 72 & \nodata & \nodata & 51 & 91 \\
 Si I & 6142.49 & 5.62 & -1.54 & 2 & 28 & \nodata & 41 & \nodata & \nodata & 38 & 41 & \nodata & 75 & 31 & 42 & 41 & \nodata & 43 & 40 \\
 Si I & 6145.02 & 5.61 & -1.48 & 2 & 35 & \nodata & 53 & 28 & 49 & 43 & 47 & \nodata & 69 & 40 & 51 & 45 & 52 & 33 & 43 \\
 Si I & 6155.14 & 5.62 & -0.84 & 1 & 75 & \nodata & \nodata & 63 & \nodata & \nodata & 73 & \nodata & 80 & 77 & 103 & 78 & 89 & 65 & 88 \\
 
\enddata

\tablecomments{References for log gf data are as follows:  (1) Thevenin (1990), (2) Reddy et al. (2003), and (3) Kurucz \& Bell (1995).  A full version of the data table can be made available upon email request.}

\end{deluxetable}

\clearpage
\setlength{\voffset}{0cm}
\setlength{\tabcolsep}{6pt}

\tablecolumns{9}
\tablewidth{0pt}

\begin{deluxetable}{lllllllll}
\tablecaption{Atomic Line Data and Equivalent Widths for NGC 2232 \label{tab7}}

\tablehead{
\colhead{Element}      &
\colhead{$\lambda$}      &
\colhead{LEP}      &
\colhead{log gf}      &
\colhead{Ref}      &
\colhead{9242}      &
\colhead{18569}      &
\colhead{18585}      &
\colhead{18588}      \\

\colhead{}      &
\colhead{($\AA$)}      &
\colhead{(eV)}      &
\colhead{}      &
\colhead{}  &
\colhead{ (m$\AA$) } &
\colhead{ (m$\AA$) } &
\colhead{ (m$\AA$) } &
\colhead{ (m$\AA$) } 

}

\startdata

  Li I\tablenotemark{a} & 6707.8 & \nodata & \nodata & 4 & 55 & 109 & 203 & 188 \\
 Al I\tablenotemark{a} & 6696.03 & 3.14 & -1.65 & 1 & 20 & 45 & 63 & 52 \\
 Al I\tablenotemark{a} & 6698.67 & 3.14 & -1.95 & 1 & 12 & \nodata & 27 & 29 \\
 Si I & 6583.71 & 5.95 & -1.67 & 1 & \nodata & \nodata & 21 & \nodata \\
 
\enddata

\tablecomments{References for log gf data are as follows:  (1) Thevenin (1990), (2) Reddy et al. (2003), (3) Kurucz \& Bell (1995), and (4) C. Sneden, private communication.}

\tablenotetext{a}{Blended feature.  Equivalent width measurements include contamination from neighboring species.  See text.  A full version of the data table can be made available upon email request.}

\end{deluxetable}

\clearpage
\setlength{\voffset}{2.2cm}
\setlength{\tabcolsep}{0.05in} 
\tablecolumns{34}
\tablewidth{0pt}

\begin{deluxetable}{llllllllllllllllllllllllllllllllll}
\rotate
\tabletypesize{\scriptsize}
\tablecaption{Stellar Abundances and Atmospheric Parameters of Stars in NGC 7160 \label{tab8}}

\tablehead{
\colhead{Star}      &
\colhead{T$_{eff}$ (K)}      &
\colhead{v$_{t}$ (km s$^{-1}$)}      &
\colhead{[FeI/H]}      &
\colhead{$\sigma$}      &
\colhead{N}      &
\colhead{[FeII/H]}      &
\colhead{$\sigma$}      &
\colhead{N}      &
\colhead{[Na/Fe]}      &
\colhead{$\sigma$}      &
\colhead{N}      &
\colhead{[SiI/Fe]}      &
\colhead{$\sigma$}      &
\colhead{N}      &
\colhead{[SiII/Fe]}      &
\colhead{$\sigma$}      &
\colhead{N}      &
\colhead{[Ca/Fe]}      &
\colhead{$\sigma$}      &
\colhead{N}      &
\colhead{[ScII/Fe]}      &
\colhead{N}      &
\colhead{[Ti/Fe]}      &
\colhead{$\sigma$}      &
\colhead{N}      &
\colhead{[V/Fe]}      &
\colhead{$\sigma$}      &
\colhead{N}      &
\colhead{[Cr/Fe]}      &
\colhead{N}      &
\colhead{[Ni/Fe]}      &
\colhead{$\sigma$}      &
\colhead{N}
        
}

\startdata

DG-40 & 6250  & 1.62  & 0.00  & 0.11  & 23  & 0.21  & 0.10  & 4 & -0.19 & \nodata & 1 & -0.03 & 0.06  & 6 & 0.13  & 0.08  & 2 & 0.02  & 0.13  & 4 & 0.13  & 1 & 0.00  & 0.07  & 3 & 0.33  & 0.31  & 2 & -0.02 & 1 & -0.05 & 0.16  & 8 \\
DG-55 & 6750  & 1.78  & 0.23  & 0.33  & 13  & \nodata & \nodata & \nodata & \nodata & \nodata & \nodata & \nodata & \nodata & \nodata & \nodata & \nodata & \nodata & \nodata & \nodata & \nodata & \nodata & \nodata & \nodata & \nodata & \nodata & \nodata & \nodata & \nodata & \nodata & \nodata & \nodata & \nodata & \nodata \\
DG-61 & 6170  & 1.59  & 0.19  & 0.14  & 22  & 0.16  & 0.12  & 4 & 0.00  & 0.18  & 2 & 0.03  & 0.08  & 5 & -0.02 & 0.02  & 2 & -0.04 & 0.07  & 4 & -0.17 & 1 & -0.08 & 0.16  & 3 & 0.13  & 0.23  & 3 & -0.15 & 1 & -0.02 & 0.12  & 9 \\

\enddata

\tablecomments{Surface gravities were set at log g = 4.1 for all stars, see text.  A full version of the data table can be made available upon email request.}

\end{deluxetable}

\clearpage
\setlength{\voffset}{2.2cm}
\setlength{\tabcolsep}{0.05in} 
\tablecolumns{16}
\tablewidth{0pt}

\begin{deluxetable}{lccccccccccccccc}

\tabletypesize{\footnotesize}
\tablecaption{Stellar Abundances and Atmospheric Parameters of Stars in NGC 2232 \label{tab9}}

\tablehead{
\colhead{Star}      &
\colhead{T$_{eff}$ (K)}      &
\colhead{v$_{t}$ (km s$^{-1}$)}      &
\colhead{[Fe/H]}      &
\colhead{$\sigma$}      &
\colhead{N}      &
\colhead{log $\epsilon$ (Li)}      &
\colhead{[Al/Fe]}      &
\colhead{$\sigma$}      &
\colhead{N}      &
\colhead{[Si/Fe]}      &
\colhead{$\sigma$}      &
\colhead{N}      &
\colhead{[Ni/Fe]}      &
\colhead{$\sigma$}      &
\colhead{N}

}

\startdata
\hline
\multicolumn{16}{c}{Abundances derived from T$_{eff}$ using (V-K)$_{0}$} \\
\hline

9242  & 6360  & 1.44  & -0.29 & 0.23  & 17  & 2.44  & 0.02  & 0.07  & 2 & -0.01 & 0.00  & 2 & -0.17 & 0.12  & 3 \\
18569 & 6600  & 1.52  & 0.30  & 0.26  & 13  & 2.99  & -0.14 & 0.00  & 1 & -0.04 & 0.00  & 1 & 0.08  & 0.29  & 3 \\
                                                              
\hline
\multicolumn{16}{c}{Abundances derived from T$_{eff}$ using SED-fitting} \\
\hline

9242  & 6500  & 1.48  & -0.21 & 0.23  & 17  & 2.54  & -0.02 & 0.08  & 2 & -0.05 & 0.00  & 2 & -0.16 & 0.12  & 3 \\
18569 & 6750  & 1.57  & 0.38  & 0.27  & 13  & 3.10  & -0.18 & 0.00  & 1 & -0.08 & 0.00  & 1 & 0.10  & 0.29  & 3 \\
\enddata

\tablecomments{Surface gravities were set at log g = 4.44 for all stars, see text.  A full version of the data table can be made available upon email request.}

\end{deluxetable}

\clearpage
\setlength{\voffset}{0cm}
\setlength{\tabcolsep}{6pt}
\tablecolumns{5}
\tablewidth{0pt}

\begin{deluxetable}{lllll}
\tabletypesize{\footnotesize}
\tablecaption{Uncertainties due to Atmospheric Parameters \label{tab10}}

\tablehead{
\colhead{Elem}      &
\colhead{T$_{eff}$}      &
\colhead{log g}      &
\colhead{v$_{t}$}      &
\colhead{Total}      \\

\colhead{}      &
\colhead{+100 K}      &
\colhead{+0.3 dex}      &
\colhead{+0.3 km s$^{-1}$}      &
\colhead{} 

}

\startdata
Fe  I & 0.08  & -0.04 & -0.07 & 0.12  \\
Fe  II  & -0.06 & 0.13  & -0.05 & 0.15  \\
Li  I & 0.09  & -0.01 & 0.00  & 0.09  \\
Na  I & 0.05  & -0.05 & -0.02 & 0.07  \\
Al  I & 0.07  & -0.01 & -0.01 & 0.07  \\
Si  I & 0.01  & 0.00  & -0.01 & 0.02  \\
Si  II  & -0.11 & 0.12  & -0.02 & 0.16  \\
Ca  I & 0.06  & -0.10 & -0.07 & 0.14  \\
Sc  II  & -0.02 & 0.12  & -0.04 & 0.13  \\
Ti  I & 0.10  & -0.04 & -0.10 & 0.15  \\
V I & 0.12  & 0.00  & -0.03 & 0.12  \\
Cr  I & 0.10  & -0.01 & -0.04 & 0.11  \\
Ni  I & 0.06  & 0.00  & -0.06 & 0.08  \\

\enddata

\end{deluxetable}

\clearpage
\tablecolumns{11}
\tablewidth{0pt}

\begin{deluxetable}{lrcllrclrcl}
\tabletypesize{\footnotesize}
\tablecaption{Mean Cluster Abundances for NGC 7160 and NGC 2232 \label{tab11}}

\tablehead{
\colhead{Element}      &
\multicolumn{3}{c}{NGC 7160}    &
\colhead{} &
\multicolumn{6}{c}{NGC 2232}      \\

\cline{2-4}
\cline{6-11} 

\colhead{} &
\colhead{} &
\colhead{} &
\colhead{} &
\colhead{} &
\multicolumn{3}{c}{T$_{eff}$ from (V-K)$_{0}$}    &
\multicolumn{3}{c}{T$_{eff}$ from SED}   

}

\startdata

$[$Fe/H]  & 0.16  & $\pm$ & 0.03  &   & 0.22  & $\pm$ & 0.09  & 0.32  & $\pm$ & 0.08  \\
$[$Fe II/H] & 0.04  & $\pm$ & 0.04  &   & \nodata &   & \nodata & \nodata &   & \nodata \\
$[$Na/Fe] & -0.07 & $\pm$ & 0.05  &   & \nodata &   & \nodata & \nodata &   & \nodata \\
$[$Al/Fe] & \nodata &   & \nodata &   & -0.22 & $\pm$ & 0.06  & -0.28 & $\pm$ & 0.06  \\
$[$Si I/Fe] & -0.06 & $\pm$ & 0.04  &   & -0.11 & $\pm$ & 0.06  & -0.19 & $\pm$ & 0.06  \\
$[$Si II/Fe]  & -0.19 & $\pm$ & 0.07  &   & \nodata &   & \nodata & \nodata &   & \nodata \\
$[$Ca/Fe] & 0.02  & $\pm$ & 0.03  &   & \nodata &   & \nodata & \nodata &   & \nodata \\
$[$Sc II/Fe]  & -0.07 & $\pm$ & 0.04  &   & \nodata &   & \nodata & \nodata &   & \nodata \\
$[$Ti/Fe] & 0.05  & $\pm$ & 0.05  &   & \nodata &   & \nodata & \nodata &   & \nodata \\
$[$V/Fe]  & 0.17  & $\pm$ & 0.04  &   & \nodata &   & \nodata & \nodata &   & \nodata \\
$[$Cr/Fe] & 0.05  & $\pm$ & 0.05  &   & \nodata &   & \nodata & \nodata &   & \nodata \\
$[$Ni/Fe] & -0.03 & $\pm$ & 0.02  &   & -0.06 & $\pm$ & 0.07  & -0.06 & $\pm$ & 0.08  \\ 
\enddata

\tablecomments{Solar abundances adopted by MOOG for this differential analysis are from Anders \& Grevesse (1989).}

\end{deluxetable}

\end{document}